\documentclass[iop]{emulateapj}
\usepackage{graphicx}

\begin{document}

\title{The origin of the most iron-poor star}
\author{S. Marassi\altaffilmark{1}}
\affil{INAF/Osservatorio Astronomico di Roma, Via di Frascati 33, 00040 Monteporzio, Italy}
\email{stefania.marassi@oa-roma.inaf.it}
\author{G. Chiaki\altaffilmark{2}}
\affil{Department of Physics, Graduate School of Science, The University of Tokyo, 
7-3-1 Hongo, Bunkyo, Tokyo 113-0033, Japan}
\author{R. Schneider\altaffilmark{1}}
\affil{INAF/Osservatorio Astronomico di Roma, Via di Frascati 33, 00040 Monteporzio, Italy}
\author{M. Limongi\altaffilmark{1}}
\affil{INAF/Osservatorio Astronomico di Roma, Via di Frascati 33, 00040 Monteporzio, Italy,
Kavli Institute for the Physics and Mathematics of the Universe, Todai Institutes for Advanced Study, 
The University of Tokyo, Kashiwa 277-8583, Japan}
\author{K. Omukai\altaffilmark{3}}
\affil{Astronomical Institute, Tohoku University, Sendai 980-8578, Japan}
\author{T. Nozawa\altaffilmark{4}}
\affil{National Astronomical Observatory of Japan, Mitaka, Tokyo 181-8588, Japan}
\author{A. Chieffi\altaffilmark{5}}
\affil{INAF/IASF, Via Fosso del Cavaliere 100, 00133 Roma, Italy}
\and
\author{N. Yoshida\altaffilmark{2,4}}
\affil{Department of Physics, Graduate School of Science, The University of Tokyo, 
7-3-1 Hongo, Bunkyo, Tokyo 113-0033, Japan}

\begin{abstract}
We investigate the origin of carbon-enhanced metal-poor (CEMP) 
stars starting from the recently discovered $\rm [Fe/H]<-7.1$ 
star SMSS J031300 (Keller et al. 2014). We show that the elemental 
abundances observed on the surface of SMSS J031300 can be well 
fit by the yields of faint, metal free, supernovae. Using properly 
calibrated faint supernova explosion models, we study, for the first time, 
the formation of dust grains in such carbon-rich, iron-poor supernova ejecta. 
Calculations are performed assuming both unmixed and uniformly mixed ejecta and 
taking into account the partial destruction by the supernova reverse shock. We 
find that, due to the paucity of refractory elements beside carbon,
amorphous carbon is the only grain species to form, with carbon condensation 
efficiencies that range between (0.15-0.84), resulting in dust yields in the 
range (0.025-2.25)M$_{\odot}$.
We follow the collapse and fragmentation of a star forming cloud enriched by 
the products of these faint supernova explosions and we explore the role 
played by fine structure line cooling and dust cooling. We show that even if
grain growth during the collapse has a minor effect of the dust-to-gas ratio,
due to C depletion into CO molecules at an early stage of the collapse, the 
formation of CEMP low-mass stars, such as SMSS J031300, could be triggered 
by dust cooling and fragmentation. A comparison between model predictions 
and observations of a sample of C-normal and C-rich metal-poor stars supports 
the idea that a single common pathway may be responsible for the formation of 
the first low-mass stars. 
\end{abstract}

\section{Introduction}
Stellar archaeology of the most metal-poor stars observed in the halo of our Galaxy
provides valuable insights on the star formation and chemical enrichment histories at high redshifts. 
The observed surface elemental abundances and the metallicity distribution function of
Galactic halo stars shed light on the nature and properties 
of the first supernovae (Pop III SNe) that pollute the interstellar medium (ISM) 
with the first metals and dust grains.

Galactic halo metal-poor stars have a well defined abundance pattern 
at $\rm [Fe/H] < -2.5$ \citep{Cay2004}that can be reasonably well reproduced by
theoretical yields of Pop III core-collapse SNe with progenitor masses in the
range $\rm 10 - 100 \, M_\odot$ \citep{Heg2010,Lim2012}.
An exception are the so-called carbon-enhanced extremely metal-poor 
stars (CEMP, defined as those with $\rm [C/Fe] > + 1$,\citep{Bee2005}, 
which comprise a fraction of $\rm \sim 10 - 20 \%$ of metal-poor stars at 
$\rm [Fe/H] <-2$ \citep{Yon2013b}). Interestingly, the observed fraction of 
C-rich metal-poor stars increases with decreasing iron abundance and,
at $\rm [Fe/H] < -4.5$, 4 out of the 5 stars currently known 
are CEMP, including the recently discovered 
SMSS J031300.36-670839.3 (hereafter SMSS J031300) with [Fe/H] $ < - 7.1$ \citep{Kel2014}.

A number of theoretical scenarios have been 
proposed to explain the origin of CEMP stars, invoking mass transfer in Pop III binary systems
\citep{Sud2004,Mas2010,Cam2010}, formation from material enriched in stellar winds of rotating 
massive and intermediate mass Pop III stars \citep{Mey2006,Mey2010,Hir2007}, or in the ejecta of 
Pop III SNe in which mixing was minimal and fallback was large. In the latter model,
the small iron abundance thus reflects the fact that in these so-called "faint" SNe
iron was made, but failed to be mixed sufficiently far out to be ejected 
\citep{Ume2003}. Alternatively, the large C/Fe could result from multiple 
generations of SNe, where the ejecta of a "normal" SN is combined with that from
a low-energy SN in which only the outer layers are ejected \citep{Lim2003},
or from a jet-induced SN explosion \citep{Tom2007}.  

The understanding of the origin of CEMP stars is complicated by the fact that
these do not form a homogeneous group and show various heavy-element
abundance patterns: depending on the level of enrichment of neutron
capture elements formed by the "s" or "r" processes, these fall in different sub-classes 
(CEMP-no, CEMP-rs, CEMP-s,\citep{Bee2005}). Generally, CEMP-s and
CEMP-rs stars have abundance patterns that suggest mass transfer from companion
AGB stars \citep{Mas2010}; indeed, most of the stars falling in these sub-classes
are members of binary systems \citep{Luc2005}. 
However, at $\rm [Fe/H] < -3$, 90\% of CEMP are CEMP-no stars, which 
do not show large enhancements of heavy neutron capture elements \citep{aok2010}. 
Using a homogeneous chemical analysis of 190 metal-poor stars
\citep{Yon2013a}. Norris et al. (2013) have recently investigated a sample of CEMP-no stars with $\rm [Fe/H] <  - 3$. 
They conclude that the observed chemical abundances are best explained by models which invoke the
nucleosynthesis of Pop III SNe with mixing and fallback, of rotating massive and intermediate mass stars
and of SNe with relativistic jets. 

Similar conclusions have been drawn by Keller et al. (2014) in their discovery paper of
SMSS J031300, a Galactic halo star with only an upper limit on the iron abundance of $\rm [Fe/H]  < -7.1$. The
observed abundance pattern of this extremely iron deficient star is optimally
matched by the nucleosynthetic yields of a $\rm 60 \, M_\odot$ Pop III SN with
a small amount of ejecta mixing due to Rayleigh-Taylor instabilities, followed by extensive
fallback of material onto the black hole remnant \citep{Jog2009,Heg2010}. 
They argue against a jet-like explosion
as this would lower the [Mg/Fe] below that observed; in addition, rapidly rotating star models are also
disfavoured by the low upper limit on N compared to C \citep{Kel2014}. 

A general conclusion of the above analyses is that below [Fe/H] $\sim -3$ there appear to be 
two populations of stars, carbon-normal and carbon-rich, whose observed elemental
abundances reflect  the nucleosynthetic products of "normal"  and "faint" Pop III SNe, 
respectively. The different composition in the material out
of which these low-mass stars form have stimulated the hypothesis that their formation
relies on two different cooling channels (Norris et al. 2013): fine-structure-line cooling and dust cooling.
In particular, C-rich stars form from C-enriched material where
gas cooling and fragmentation is dominated by CII (and eventually OI) fine structure lines
\citep{Bro2003}. 

Frebel et al. (2007) have introduced the so-called transition discriminant, $\rm D_{trans}= log(10^{[C/H]} + 0.9 \times 10^{[O/H]})$
and predicted that metal-poor stars forming through this mechanism must have $\rm D_{trans} > -3.5 \pm 0.2$ \citep{Fre2013}.
Nearly all stars satisfy this criterion with the exception of SDSS J1029151+1729 (Caffau et al. 2011). 
This is a carbon-normal star with [Fe/H]$ = -4.99$ and a solar-like chemical abundance pattern, leading
to a total metallicity of $\rm Z = 4.5 \times 10^{-5} Z_{\odot}$ and representing
 the most chemically pristine object currently
known. The origin of the star has been investigated by Schneider et al. (2012b) who suggest that the observed 
abundance pattern of  SDSS J1029151 can be well matched by yields of "normal" Pop III SNe with $\rm 20$ and $\rm 35 \, M_{\odot}$;
the mass of dust formed in these SN ejecta is enough to satisfy the minimal conditions to activate dust-induced fragmentation,
which require the gas to be pre-enriched above a critical dust-to-gas mass ratio ${\cal D}_{\rm cr} =  [2.6 - 6.3] \times 10^{-9}$,
with the spread reflecting the dependence on the grain properties \citep{Sch2003,Sch2010,Sch2012a}.
Note that even if dust formation in the first SNe is less efficient or if strong dust destruction by the SN reverse shock occurs, 
grain growth during the collapse of the parent gas cloud is sufficiently rapid to activate dust cooling and fragmentation 
\citep{Noz2012,Chi2013}.

In principle, the values ${\cal D}_{\rm cr} = [2.6 - 6.3] \times 10^{-9}$ and $\rm D_{trans} > -3.5 \pm 0.2$
could provide a way to test dust cooling and fine-structure-line cooling against the observations. This
has been recently explored by Ji et al. (2014) who - however - make the apriori assumption
that dust is made by silicates. In this way they estimate the critical silicate abundance which corresponds to 
${\cal D}_{\rm cr}$ and compare these values to chemical abundances of metal-poor stars, with [Fe/H] $ < - 4$. 
They find that the stars exhibit either high carbon with low silicon abundances or the reverse. 
A silicate dust scenario would thus support the hypothesis that the earliest low-mass star formation 
in the most metal-poor regime may have proceeded by means of two distinct cooling pathways: 
fine-structure line cooling and dust cooling.

Here we intend to investigate the origin of CEMP-no stars, and in particular of the
recently discovered most iron-deficient member of this class, SMSS J031300
\citep{Kel2014}. For the first time, we consider the formation of dust 
in faint Pop III SN ejecta. Our aim is to verify if dust can form in these C-rich, iron-poor ejecta, 
and what is the relative role of fine-structure-line cooling and dust cooling in the
collapse of material enriched by these explosions.

Having this goal in mind, we first compare the observed abundances of SMSS J031300
with the nucleosynthetic yields of model Pop III SNe that range in progenitor mass, 
internal mixing, and explosion energy (extent of fallback, Section 2). 
We perform dust calculation in the ejecta of the selected SN models with a revised version of 
the Bianchi \& Schneider (2007) model which includes an improved, non-steady state, 
molecular network, updated radiative association reactions 
for CO, SiO, C$_2$ and O$_2$ molecules, and formation/destruction 
processes between molecules (Marassi et al. 2014a; Marassi et al. 2014b, Section 3). We follow the collapse of a gas cloud 
assuming that the newly formed dust and the metals present in the ejecta 
are uniformly mixed into the collapsing cloud with the same total metallicity 
that is observed on the surface of SMSS J031300 and verify under which 
conditions the formation of low-mass star is enabled (Section 4). 
Finally, we discuss the implications of our results for the origin of the first
low-mass stars in the Universe (Section 5). 

\begin{figure*}[t!]
\hspace{-1.0cm}
\includegraphics[angle=0,width=11.0cm]{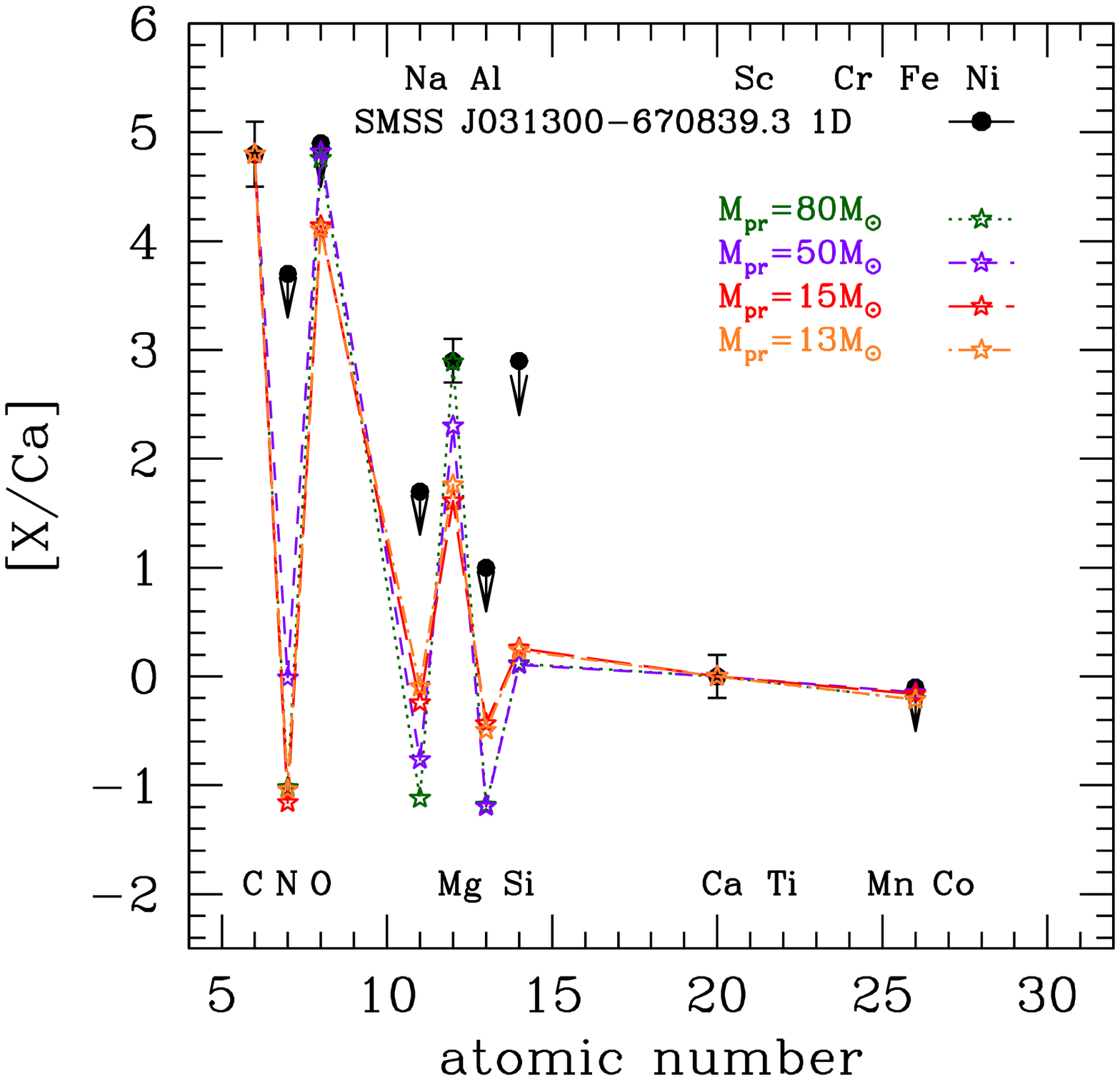}
\hspace{-3.45cm}
\includegraphics[angle=0,width=11.0cm]{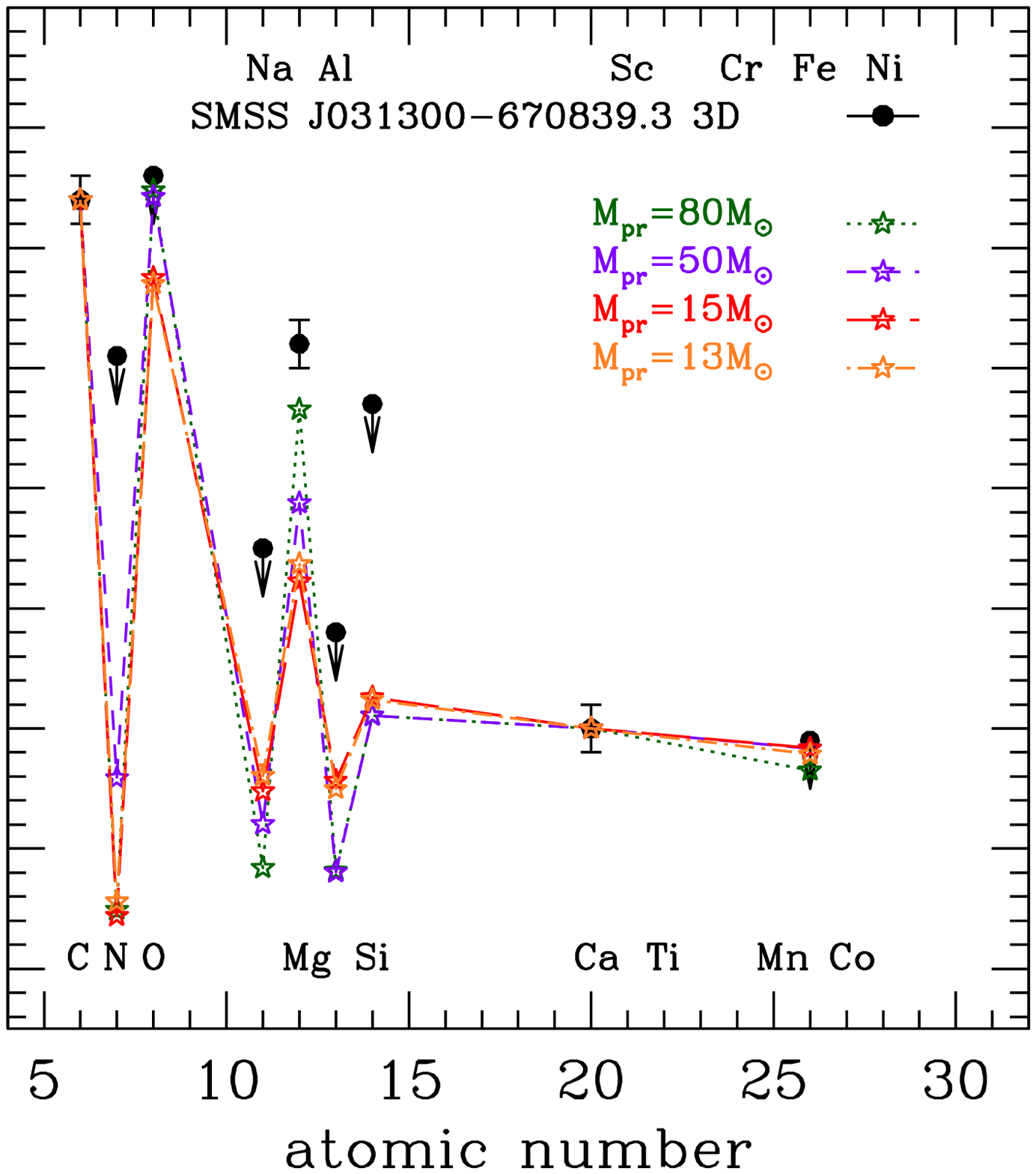}
\caption{Comparison between the observed element abundance ratios of
the star  SMSS J031300 and the chemical yields of Pop III faint SN 
with progenitor masses of $\rm 13, 15, 50, and \,80~M_{\odot}$. Mixing and fallback are chosen so as to minimize the scatter with 
the observations using a 1D (left panel) or
3D (right panel) model atmosphere (black points). Dots with arrows show upper limits and filled points with 
errorbars indicate the detections, where an error of 0.1 (0.2) has been associated to 
the observed Ca and Mg (C) abundances.}
\label{fig:fit1D3D}
\end{figure*}

\section{Faint supernova progenitors}

Our aim here is to identify a set of plausible Pop III
SN models which may have polluted the birth cloud of SMSS J031300.
The procedure that we follow is to minimize the scatter with the observed abundance pattern 
adopting the chemical yields produced by a set of Pop III core-collapse SN models
with mixing and fallback.

We refer the interested reader to Limongi \& Chieffi (2012) for a thorough presentation
of the models. Similarly to what has been done in Schneider et al. (2012b), we consider a set 
of models that extends in mass between 13 and 80 $\rm M_\odot$, which appear to
be consistent with the mass range found by
the most recent simulations of first star formation (Hirano et al. 2014). These models 
have been followed from the pre-main sequence phase up to the onset of the iron core collapse by 
means of the FRANEC stellar evolutionary code. The initial composition is set to the 
pristine Big Bang nucleosynthesis (Y=0.23) and mass loss is switched off during 
all the evolutionary stages. 

The explosive nucleosynthesis has been computed in the framework of the induced explosion 
\citep{Lim2006} approximation. For each progenitor model, we identify 
a combination of mixing and fallback
that provides the best-fit to the observed elemental abundance ratios (or
upper limits) on the surface of SMSS J031300. The observational data 
points are shown in Fig.\ref{fig:fit1D3D}: black points with arrows show the 
inferred 1D/3D model atmosphere upper limits and the filled black dots with errorbars are
the 1D/3D model atmosphere detections (see Keller et al. 2014 for a description of the
model atmosphere used for the abundance analysis). Among the three elements that have been detected -
C, Mg, and Ca - we show the elemental abundances relative to Ca. 

The low upper limit on [N/Ca] disfavour
SN progenitors with initial masses in the range $\rm [20 - 35] \, M_{\odot}$.
In fact, their evolution leads to comparable C and N abundances in 
the ejecta (see Fig.7 and the discussion in Limongi \& Chieffi 2012), in contrast to the observed 
pattern in SMSS J031300. Hence, we include in the analysis SN progenitors 
with initial masses equal to $\rm 13, 15, 50, and \, 80 \, \rm M_\odot$. For each of
these models, we have applied the optimal combination of mixing (set
by a minimum and maximum mass coordinates within which the mass
of each element is uniformly mixed in the ejecta) and fallback (set by
the mass-cut) that minimizes the scatter with the observed abundance
pattern. Since the elemental abundances inferred from the analysis 
using 1D or 3D model atmospheres lead to sensible differences in
the observed abundances (or upper limits) of elements that are critical for the 
resulting dust mass, such as C, O, and Mg, for each progenitor mass 
we have independently inferred best-fit models using 1D and 3D observed 
elemental abundances. The resulting models are shown by the starred points joined
with dotted lines in Fig.\ref{fig:fit1D3D}.
It is clear from the figure that while all the models show an overall agreement with the observed
abudance pattern, it is hard to simultaneously accomodate the upper limit on [O/Ca] and the
detected [Mg/Ca]; this is particularly true for low mass progenitors, whose evolution leads to
a lower [Mg/O] in the ejecta. The best-fit model appears to be the most massive progenitor
with $\rm 80 \, M_\odot$: adopting the abundance pattern inferred from 1D atmosphere models,
the predicted ejecta composition is compatible with all the data points. This is not true, however,
when the 3D data points are considered (see the right panel in Fig.\ref{fig:fit1D3D}) as the
more stringent upper limit on [O/Ca] and the larger [Mg/Ca] cannot be reproduced. 
Note that in Keller et al. (2014), the model that provides the optimal match to the observed abundance
pattern is a $1.8 \times 10^{51} \rm erg$ explosion of a $60 \rm \, M_\odot$ star of primordial
composition with a small amount of ejecta mixing and a strong fallback. These properties closely
resemble our best-fit model and indeed the predicted [Mg/H] falls below the observational data point
(see Fig. 3 in Keller et al. 2014). 

Table  \ref{tab:masses} summarizes the main properties of the SN progenitors. 
For all models, we choose the interval of mixing and mass-cut to reproduce the observed
[C/Ca] and maximize the [Mg/Ca], getting as close as possible to the observed value without
exceeding the [O/Ca] and [Fe/Ca] upper limits. We find that for all but the most massive model, 
the ejecta composition is insensitive to the reference set of observed abundances, whether
derived with 1D or 3D model atmosphere (the values quoted in the Table refer to the 3D set). 
Conversely, the chemical structure of the  $\rm 80 \, M_\odot$ progenitor, characterized by a
more extended Mg shell, leads to [O/Ca] and [C/Ca] abundance ratios that are more sensitive
to the mixing and fallback prescription adopted. As a result, in Table \ref{tab:masses} 
we show the ejecta composition for both sets of observed abundances and refer to the models
as $\rm 80 M_\odot$-1D or $\rm 80 M_\odot$-3D. 
\begin{table*}
\caption{Properties of faint SN progenitors, including the explosion energy [$10^{51}$~erg], the ejecta velocity [$\rm km~s^{-1}$], 
the gas number density [$\rm cm^{-3}$] and the radius of He core, $\rm R_0 $ [cm] at $\rm t=t_{0}$ [sec] when adiabatic expansion starts (see text);
in addition, we also show  the helium core mass, the mass of the ejecta, and the initial masses of the elemental yields that are observed on the surface of SMSS J031300, 
C, N, O, Na, Mg, Al, Si, Ca and Fe [$\rm M_{\odot}$].
Each model name identifies the progenitor mass; the additional labels - where present - 1D or 3D specify the set of observed abundances used in model calibration 
(with default values being 3D, see text). The last two columns identify the two layers (A and B) of the unmixed ejecta of the $\rm 80 M_\odot$-1D model.}
\label{tab:masses}
\begin{tabular}{@{}lccccccc}
\hline
       & $\rm 13 M_{\odot}$ &  $\rm 15 M_\odot$ & $\rm 50 M_\odot$ & $\rm 80 M_\odot$  & $\rm 80 M_\odot$  &$\rm 80 M_\odot$ &$\rm 80 M_\odot$ \\
\hline
& 3D & 3D & 3D & 3D & 1D & 1D-A & 1D-B\\
\hline
 $\rm E_{exp} $ & $0.5$& $0.7$ & $2.6$&  $5.2$& $5.2$ &  $5.2$ & $5.2$\\
 $\rm M_{He_{core}}$ & $2.81$& $3.40$& $18.11$& $31.94$ &  $31.94$ &$31.94$&$31.94$\\
 $\rm M_{cut}$& $1.7$& $2.1$& $11$& $22.5$& $24$& $24$& $24$\\
 $\rm v_{eje}$& $2722$ & $3011$ & $3344$& $3891$& $3943$& $3943$&$3943$\\          
 $\rm M_{eje}$& 11.3 & 12.9 & 39 & 57.5 & 56 & 56 & 56 \\
\hline
$\rm R_{0} $ &$2.65\times 10^{14}$&$2.44\times 10^{14}$&$2.51\times 10^{14}$&$4.78\times 10^{14}$&$4.78\times 10^{14}$& $2.29\times 10^{14}$ &$4.57\times 10^{14}$\\
$\rm t_{0} $ &$2.11\times 10^{6}$&$1.78\times 10^{6}$&$3.12\times 10^{6}$&$4.14\times 10^{6}$&$4.14\times 10^{6}$& $2.24\times 10^{6}$& $4.14\times 10^{6}$\\
$\rm n_{0} $ &$1.14\times 10^{11}$&$2.73\times 10^{11}$&$4.37\times 10^{12}$&$7.64\times 10^{11}$& $5.17\times 10^{11}$&$2.33\times 10^{13}$&$3.31\times 10^{11}$\\
\hline
 $\rm M_{C}$ & 0.065 &  0.118  &  0.989 &   1.089 &  0.887  &  0.547  &  0.339 \\
 $\rm M_{N}$ & $2.72\times 10^{-8}$ & $3.71\times 10^{-8}$ & $4.45\times 10^{-6}$ & $3.93\times 10^{-7}$& $3.93\times 10^{-7}$&-&-\\                       
 $\rm M_{O}$ & 0.032 &  0.065  &  2.551 &   3.213 &  1.973 &  1.948  &  0.024  \\
$\rm M_{Na}$ & $1.29\times 10^{-8}$ & $1.74\times 10^{-8}$ & $7.79\times 10^{-8}$ & $3.70\times 10^{-8}$& $1.31\times 10^{-8}$& $1.31\times 10^{-8}$ &-\\
$\rm M_{Mg}$ & $1.81\times 10^{-5}$ & $2.36\times 10^{-5}$ & $8.84\times 10^{-4}$ & $5.90\times 10^{-3}$& $3.26\times 10^{-3}$& $3.26\times 10^{-3}$ &-\\ 
$\rm M_{Al}$ & $1.91\times 10^{-8}$ & $4.06\times 10^{-8}$ & $5.89\times 10^{-8}$ & $6.70\times 10^{-8}$& $2.17\times 10^{-8}$& $2.17\times 10^{-8}$ &-\\
$\rm M_{Si}$ & $1.26\times 10^{-6}$ & $2.41\times 10^{-6}$ & $1.42\times 10^{-5}$ & $1.57\times 10^{-5}$& $5.19\times 10^{-6}$& $5.19\times 10^{-6}$ &-\\
$\rm M_{Ca}$ & $7.03\times 10^{-8}$ & $1.28\times 10^{-7}$ & $1.07\times 10^{-6}$ & $1.18\times 10^{-6}$& $3.82\times 10^{-7}$& $3.82\times 10^{-7}$ &-\\
$\rm M_{Fe}$ & $8.62\times 10^{-7}$ & $1.76\times 10^{-6}$ & $1.47\times 10^{-5}$ & $1.05\times 10^{-5}$& $4.7\times 10^{-6}$ & $4.7 \times 10^{-6}$ &-\\
\hline
\end{tabular}
\end{table*}
\begin{table*}
\caption{Masses of CO, SiO, $\rm O_{2}$, $\rm C_{2}$ molecules in faint SN progenitor ejecta and masses of AC grains [$\rm M_{\odot}$]. 
Each model name identifies the progenitor mass; the additional labels - where present - 1D or 3D specify the set of observed abundances 
used in model calibration (with default values being 3D, see text). The last two columns refer to
the two layers (A and B) of the unmixed ejecta of the $\rm 80 M_\odot$-1D model.}
\label{tab:molecules}
\begin{tabular}{@{}lccccccc}
\hline
       & $\rm 13 M_{\odot}$ &  $\rm 15 M_\odot$ & $\rm 50 M_\odot$ & $\rm 80 M_\odot$  & $\rm 80 M_\odot$  &$\rm 80 M_\odot$ &$\rm 80 M_\odot$ \\
\hline
& 3D & 3D & 3D & 3D & 1D & 1D-A & 1D-B\\
\hline
 $\rm M_{CO}$  &  0.0246  & 0.074   & 2.29   &  2.15  & 1.44  & 1.276   & 0.043  \\
 $\rm M_{O_2}$ & $2.91\times 10^{-8}$& $5.77\times 10^{-8}$& $6.21\times 10^{-5}$&$6.29\times 10^{-5}$& $2.42\times 10^{-5}$& $0.513$ &- \\
 $\rm M_{C_2}$ & - & - & - & - & - & -& $0.21$ \\                            
 $\rm M_{SiO}$ & $8.60\times 10^{-8}$& $2.73\times 10^{-7}$ & $2.02\times 10^{-5}$& $1.49\times 10^{-5}$& $3.48\times 10^{-6}$& $8.11\times 10^{-6}$ &-\\
\hline
 $\rm M_{AC}$  &  0.0544   & 0.0862  & 0.00557  & 0.165   & 0.269 &  -   & 0.112\\
\hline
\end{tabular}
\end{table*}

\section{Dust formation in faint SN}

The formation of dust in SN ejecta has been modeled following  either 
classical nucleation theory (CNT, Todini \& Ferrara 2001; Nozawa \& Kozasa 2003; Nozawa et al.~2010; Bianchi \& Schneider 2007; 
and references therein) or a chemical kinetic approach \citep{CD2009,CD2010,Sar2013}. 
As recently pointed out in Cherchneff \& Dwek (2009), Cherchneff \& Dwek (2010) and Sarangi \&
Cherchneff (2013), these methods lead to different estimates of the mass and composition 
of newly formed dust due to the questionable applicability of the standard CNT in SN ejecta
(Donn \& Nuth 1985) and to non-equilibrium chemical processes related to the formation/destruction
of molecules  (Cherchneff \& Dwek 2009).  Recently, Paquette \& Nuth (2011) have shown that 
the assumption of chemical equilibrium at nucleation, adopted within CNT, 
has only a minor effect on the grain mass and size distribution. In addition, Nozawa \& Kozasa (2013)
have demonstrated that a steady-state nucleation rate is a good approximation in SN ejecta, at least
until the collisional timescales of the key molecule is much smaller than the
timescale with which the supersaturation ratio increases.

Here we follow CNT to compute the nucleation and accretion of dust species using a previously 
developed code that has been applied to core-collapse \citep{Tod2001,Bia2007} and pair-instability 
SNe \citep{Sch2004} and that can successfully reproduce observational data of SNe and young SN 
remnants \citep{Sch2014,Val2014}.

For the present study, the code has been improved allowing to solve  the non
steady-state formation of important molecular species, such as CO, O$_2$, C$_2$, and SiO. A full description
of the upgraded molecular network with the reactions rates will be given in Marassi et al. (2014a). Here
we restrict the discussion to those aspects that are relevant for carbon dust formation in
faint SN ejecta. It is important to stress that the upgraded molecular network has been
tested against the results obtained by chemical kinetic approach: in Marassi et al. (2014b) we have 
applied the model to a $\rm 15 \, M_\odot$ type-IIP SN with solar metallicity 
and a stratified ejecta, and we find that the mass of CO at the onset of dust nucleation is 
in excellent agreement with results obtained by Sarangi \& Cherchneff (2013) for the same
SN progenitor.\\

We construct the ejecta models using as initial conditions the thermo-dynamical properties 
obtained by the SN explosion simulations (Limongi \& Chieffi 2012). 
The ejecta follow an adiabatic expansion and the temperature evolution is given by, 
\begin{equation}
\rm{T}=\rm{T_{\rm 0}}\left[1+\frac{\rm v_{eje}}{R_{\rm 0}}t\right]^{3(1-\gamma)}
\end{equation}  
where $\gamma=1.41$ is the adiabatic index, $\rm v_{eje}$ is the ejecta expansion velocity, 
\begin{equation}
\rm v_{eje}=\sqrt{\frac{10 \rm E_{\rm kin}}{3\rm M_{\rm eje}}}\,
\end{equation}
and $\rm T_{0}$ and  $R_0$ are the temperature and radius of the He core at the initial
time $\rm t=t_0$ and are extracted from the simulation outputs. 
The initial time $t_0$ is chosen to correspond to the time when the radius of the He core, $\rm R_{He_{core}}$ 
reaches a temperature of $\rm T_{0}=10^{4}$\,K. All these quantities and the initial gas number density $\rm n_{0}(t=t_{\rm 0})$ 
are shown in Table 1 for different SN progenitor models.

It is well known that the formation of CO molecules subtracts C-atoms limiting 
the formation of carbon grains. In the C-rich and Fe-poor ejecta of faint SN,  
the main molecular destruction process, collisions with energetic electrons 
produced by the radioactive decay of $^{56}$Co, is strongly inhibited.

For all the faint SN models that we have explored, the formation of
CO is dominated by radiative association reaction \citep{Dal1990},
\begin{equation}
\rm C+O\rightarrow CO +h\nu \,
\label{radCO}
\end{equation}
and by bimolecular, neutral-neutral reactions \citep{Umist}
\begin{equation}
\rm C+O_{2}\rightarrow CO+O \qquad O+C_{2}\rightarrow CO+C\,
\label{nnCO}
\end{equation}
that involve O$_2$, C$_2$ molecule, and C, O atoms. 
As a result, the mass of CO that is formed ranges between 0.0544\,M$_{\odot}$ 
and 0.269\,M$_{\odot}$ (see Table 2). The formation of SiO in faint SN ejecta 
is strongly suppressed with respect to ordinary core-collapse SNe, as a consequence
of the very small initial abundance of Si (see Tables 1 and 2). Additional molecular 
species, such as $\rm O_2$ and $\rm C_2$, form through radiative association reactions \citep{AS1997,Babb1995}, 
\begin{equation}
\rm C+C\rightarrow C_2 + h\nu \qquad O+O\rightarrow O_{2} +h\nu\,
\end{equation}
and through the inverse bimolecular process showed in eq.~(\ref{nnCO}) in non-negligible 
amount only if the ejecta is stratified (see Table 2).

Following Bianchi \& Schneider (2007) our dust calculation is based on CNT. When a gas 
becomes supersaturated, particles (monomers) aggregate in a seed cluster which subsequently 
grows by accretion of other monomers. The accretion process is regulated by the collision rate 
of the key species and depends on the sticking coefficient, defined as the probability that an atom 
colliding with a grain will stick to it. We show the results obtained assuming that the 
seed clusters are formed by a minimum of 2 monomers and that the sticking coefficient is 
equal to 1 for all grain species. 
The chemical composition and size at condensation and the basic data necessary for the 
dust formation calculations are shown in Table 2 of Nozawa et al. (2003). 
The thermal and optical properties of the resulting dust materials can be found in Table A1 of 
Bianchi \& Schneider (2007), to which we refer the interested reader.
The nucleation process, together with accretion, results 
in a typical lognormal grain size distributions \citep{Tod2001,NK2003}.\\
If faint SN are believed to be powered by a spherical-like
explosion, mixing occurs by means of Rayleigh-Taylor instabilities only in the innermost 
shells \citep{Jog2009}, up to a mass-coordinate that is very close to the mass-cut 
($M_{cut}$). Hence the material beyond the mass cut is likely to be stratified and 
dust nucleation in unmixed ejecta provides a more realistic estimates of the total mass 
of dust produced. To enable a more direct comparison with previous results (Schneider et al. 2012b),
where dust formation in uniformly mixed, Pop III ordinary core-collapse SNe has been considered, 
we first present the results adopting uniformly mixed ejecta and then extend the calculation 
to stratified ejecta.\\

\subsection{Dust formation in uniformly mixed ejecta}

\begin{figure*}[t!]
\includegraphics[angle=0,width=9.5cm]{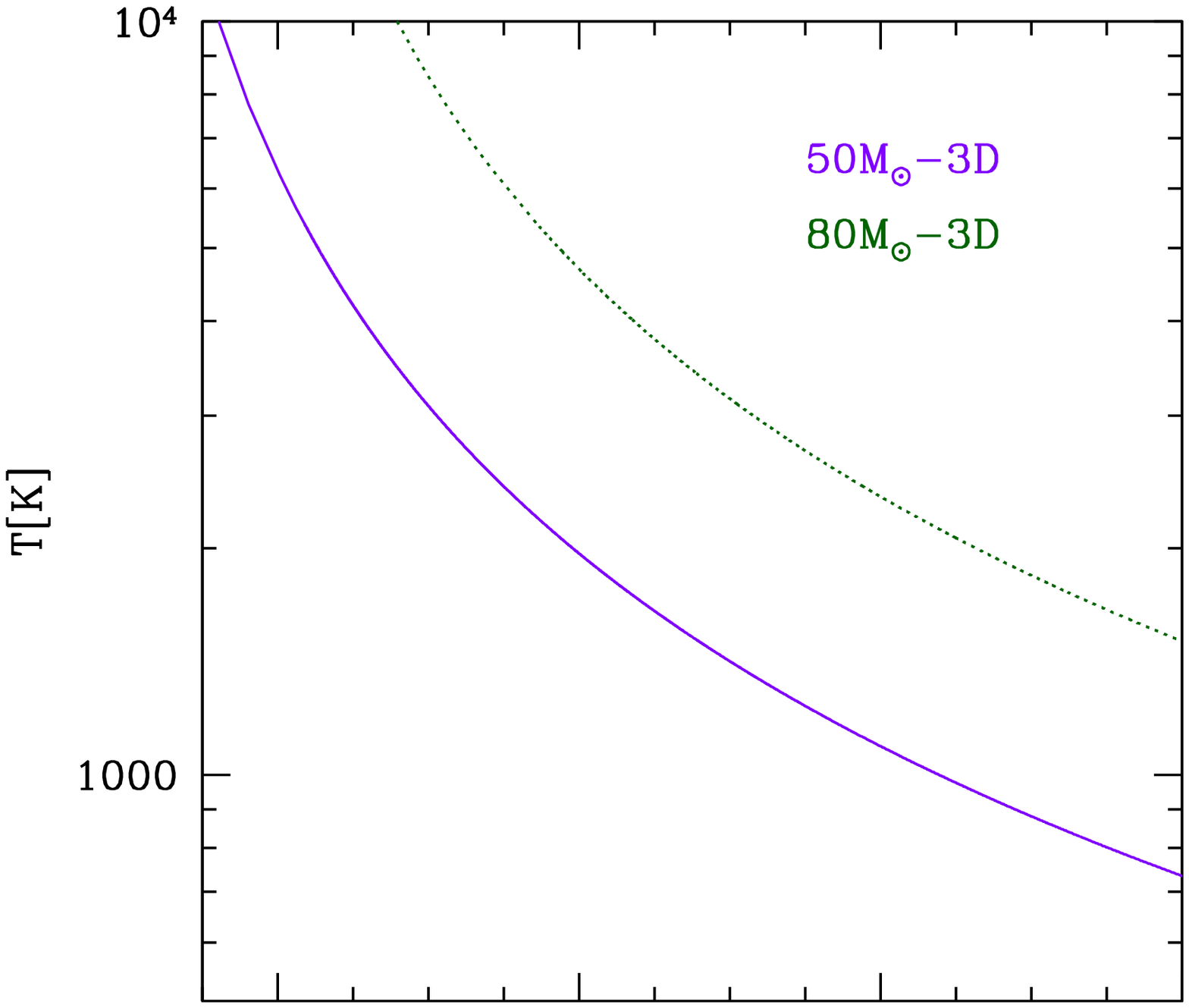}
\vspace{-2.9cm}
\hspace{-1.5cm}
\includegraphics[angle=0,width=9.5cm]{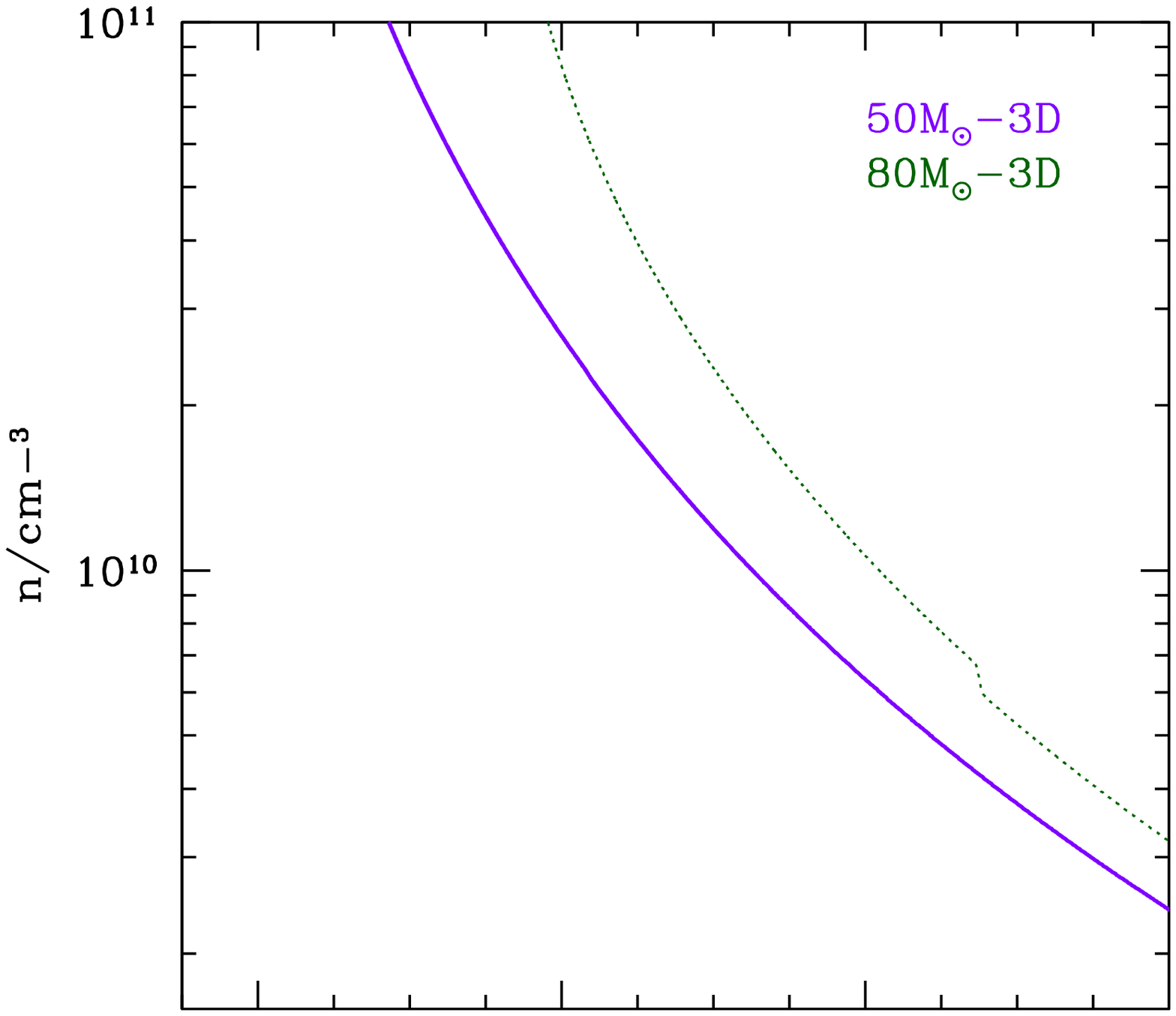}
\hspace{-2.5cm}
\includegraphics[angle=0,width=9.5cm]{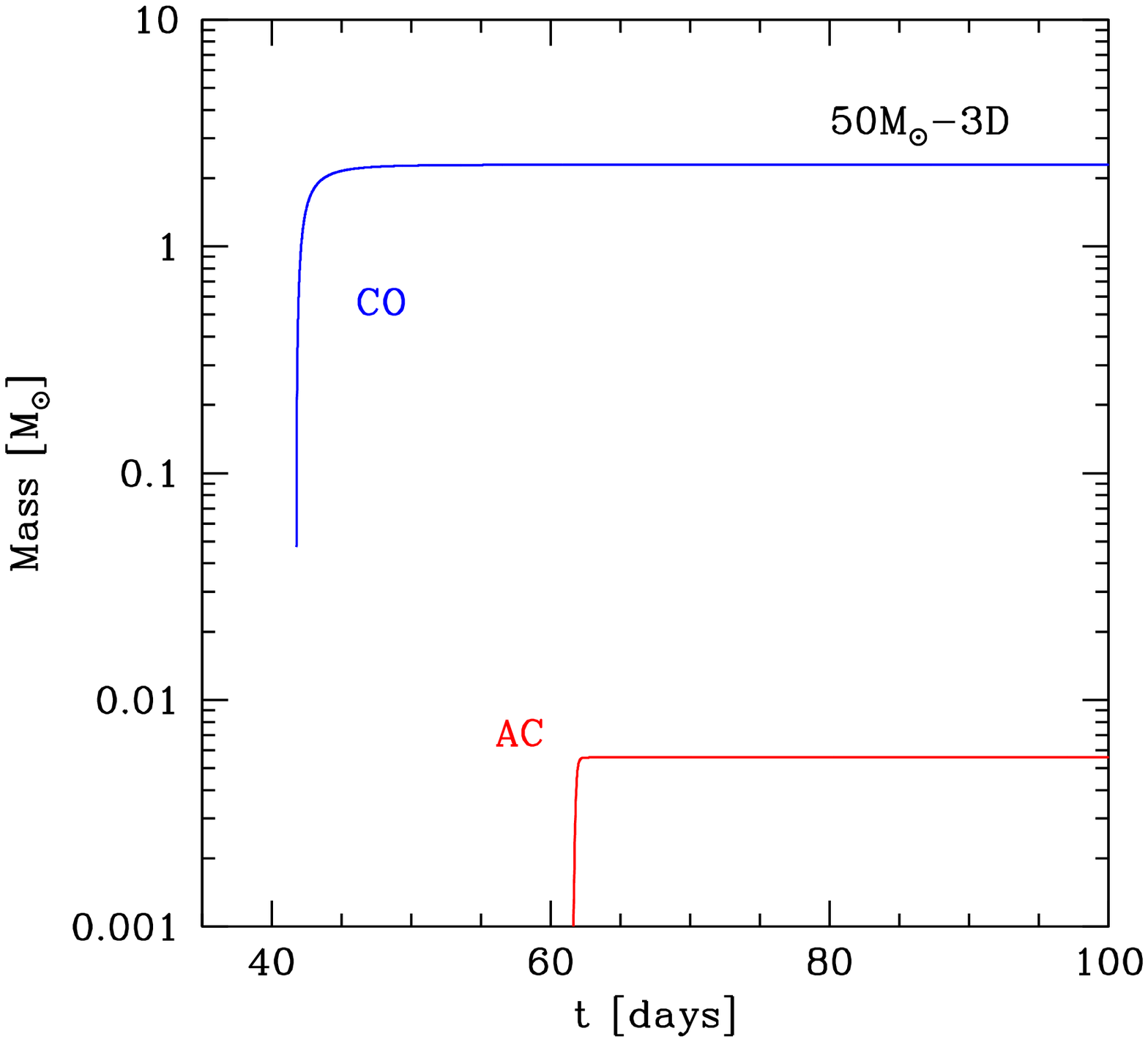}
\hspace{-1.16cm}
\includegraphics[angle=0,width=9.5cm]{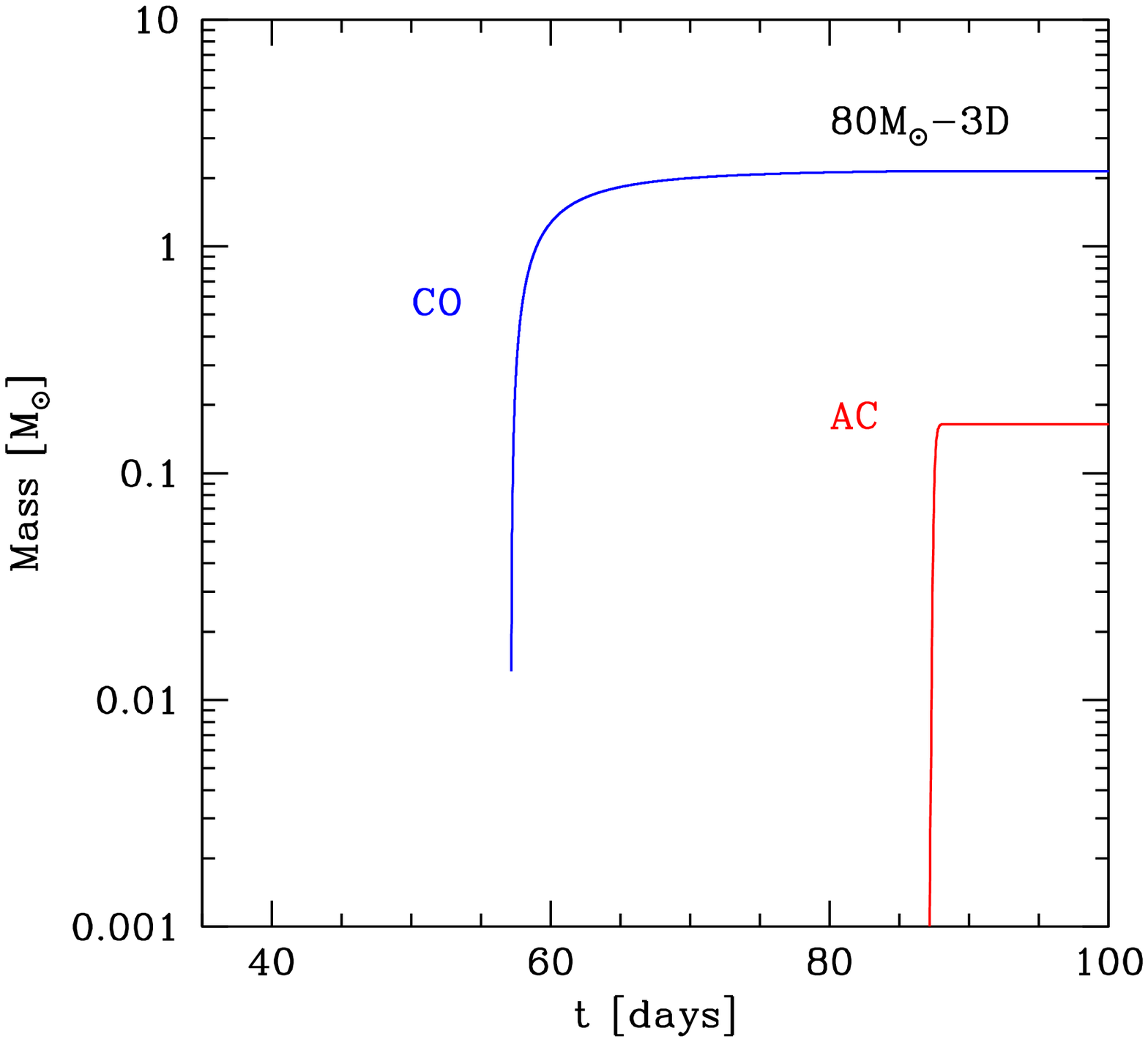}
\hspace{-1.5cm}
\caption{Upper panels: temperature and density evolution 
for $\rm 50 \, M_\odot$ and $\rm 80 \, M_\odot$ 3D mixed models. 
Bottom panels: time evolution of CO molecule and dust
for $\rm 50 \, M_\odot$ and $\rm 80 \, M_\odot$ 3D mixed models.}
\label{fig:80_50_3D}
\end{figure*}

Contrary to ordinary core collapse SN \citep{Sch2012b}, the only grain species that 
form in non negligible amount in faint SN is Amorphous Carbon (AC). This is due to 
the peculiar metal composition of faint SN ejecta, that are dominated by C-atoms and 
where the abundances of Mg, Si, Al and Fe are too small to enable the formation of 
silicates, magnetite or alumina grains.   
In Table \ref{tab:molecules} we show the resulting masses of molecules and of 
AC grains for all the explored faint SN models. 
We show that the mass of dust increases with the mass of the SN progenitor, ranging 
from $\rm M_{dust} = 5.4 \times 10^{-2} \,M_\odot$ for the 13 $\rm \,M_{\odot}$ model to  
$\rm M_{dust} = 0.269 \, (0.165) \,M_\odot$ for the 80 $\rm \,M_{\odot}$-1D (3D) model, 
consistent with the larger metal mass initially present in the ejecta. The only exception 
is represented by the $\rm 50 \, M_\odot$ model, which produces the smallest mass of dust, 
$\rm M_{dust} = 5.6 \times 10^{-3}\, M_\odot$. The reason for this behaviour resides in the 
larger mean ejecta density of the $\rm 50 \,M_\odot$ model, which increases the rates of 
the major processes leading to the formation of CO molecules. 
This can be understood by looking at  Fig.~\ref{fig:80_50_3D} where we show the time evolution of 
temperature, density, CO and AC masses for the $\rm 50 \, M_{\odot}$ and the $\rm 80 \, M_{\odot}$ models. 
In the $\rm 50 \, M_{\odot}$ SN model, dust nucleation starts when the temperature has dropped below the AC grain 
condensation temperature, $\sim 1940$\,K, roughly 60 days after the explosion, when the number density is $\rm n_{0}=2.58\times 10^{10}\rm cm^{-3}$. 
Conversely, in the $\rm 80 \, M_{\odot}$ SN model, the gas temperature drops to $\sim 2000$\,K only 86 days after the explosion,
when the number density is $\rm n_{0}=7.15\times 10^{9}\rm cm^{-3}$.
Comparing the chemical reaction rates to the expansion rate, ${\rm k}_{\rm dyn}$, defined as
\begin{equation}
\rm{k}_{\rm dyn} = 1/t = \frac{v_{\rm eje}}{R(t)}
\end{equation}
where v$_{\rm eje}$ is the ejecta velocity and R(t) is the ejecta position at a given time t after
explosion, we can determine $\rm t = t_{\rm fo}$, the time at which each reaction rate becomes
longer than the expansion rate, leading to a "freeze-out" of the corresponding chemical species 
(Cherchneff \& Dwek 2009).
The mass of CO molecules at $\rm t_{fo}$ and at the onset of dust nucleation ($\rm t_{\rm nucl}$) 
largely determines the resulting dust mass. For the $\rm 50 \, M_\odot$ model, 
we find that $\rm M_{CO}(t_{\rm fo}) = 2.21 \rm \, M_{\odot}$ and, at this stage, 
the carbon mass locked in CO molecules is $\rm 0.945 \, M_\odot$. 
When nucleation starts, $\rm M_{CO}(t_{\rm nucl}) = 2.29 \rm \,M_{\odot}$ and $\rm 0.982 \, M_\odot$ mass of carbon is
locked in CO molecules. Hence, out of the  $0.988\rm \,M_{\odot}$ of C atoms initially present in the ejecta
(see Table 1),  only $\sim 6\times 10^{-3} \rm M_{\odot}$ are available to form C-grains.
If we do the same calculation for the $\rm 80 M_\odot$ - 3D model, we find
$\rm M_{CO}(t_{\rm fo}) = 1.258 \rm \,M_{\odot}$, $\rm M_{CO}(t_{\rm nucl}) =1.438\rm \,M_{\odot}$, and the mass of carbon 
atoms locked in CO at nucleation is $0.616 \rm \,M_{\odot}$, so that there are still $0.27\rm \, M_\odot$ of carbon
free to form C-grains. Hence, the reason why the $50 \rm \,M_\odot$ model forms the least amount of dust is
the higher efficiency of CO formation which leads to a carbon depletion in CO molecules
of  $95 \%$ at freeze-out. Note that in less massive progenitor models, the ejecta 
composition is characterized by an initial mass of C greater than the O mass (see Table \ref{tab:masses});
this contributes to decrease C depletion into CO molecules, increasing the resulting mass of AC grains. 
\\

\subsection{Dust formation in unmixed ejecta}

\begin{figure*}[t!]
\includegraphics[angle=0,width=8.5cm]{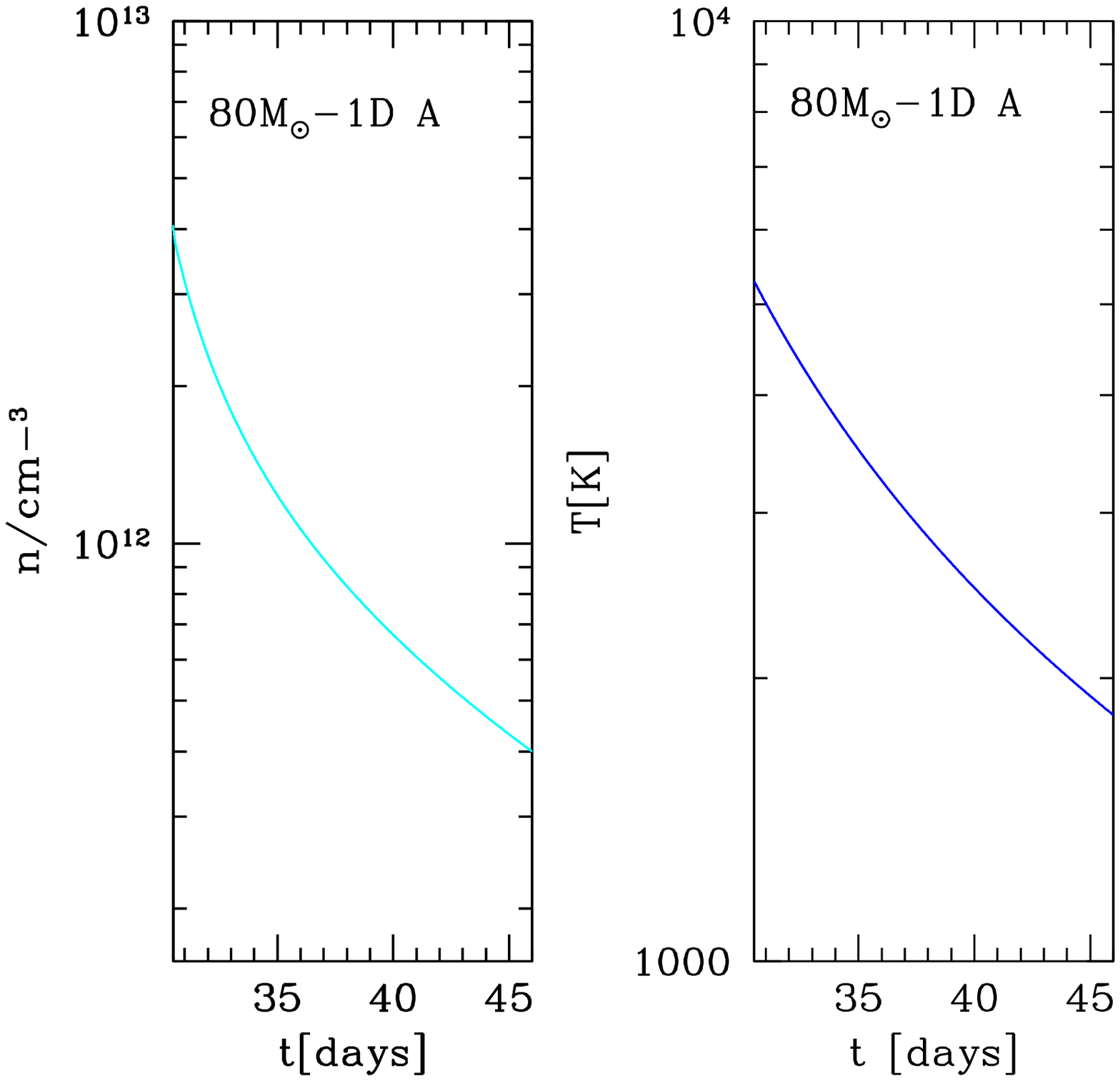}
\includegraphics[angle=0,width=8.5cm]{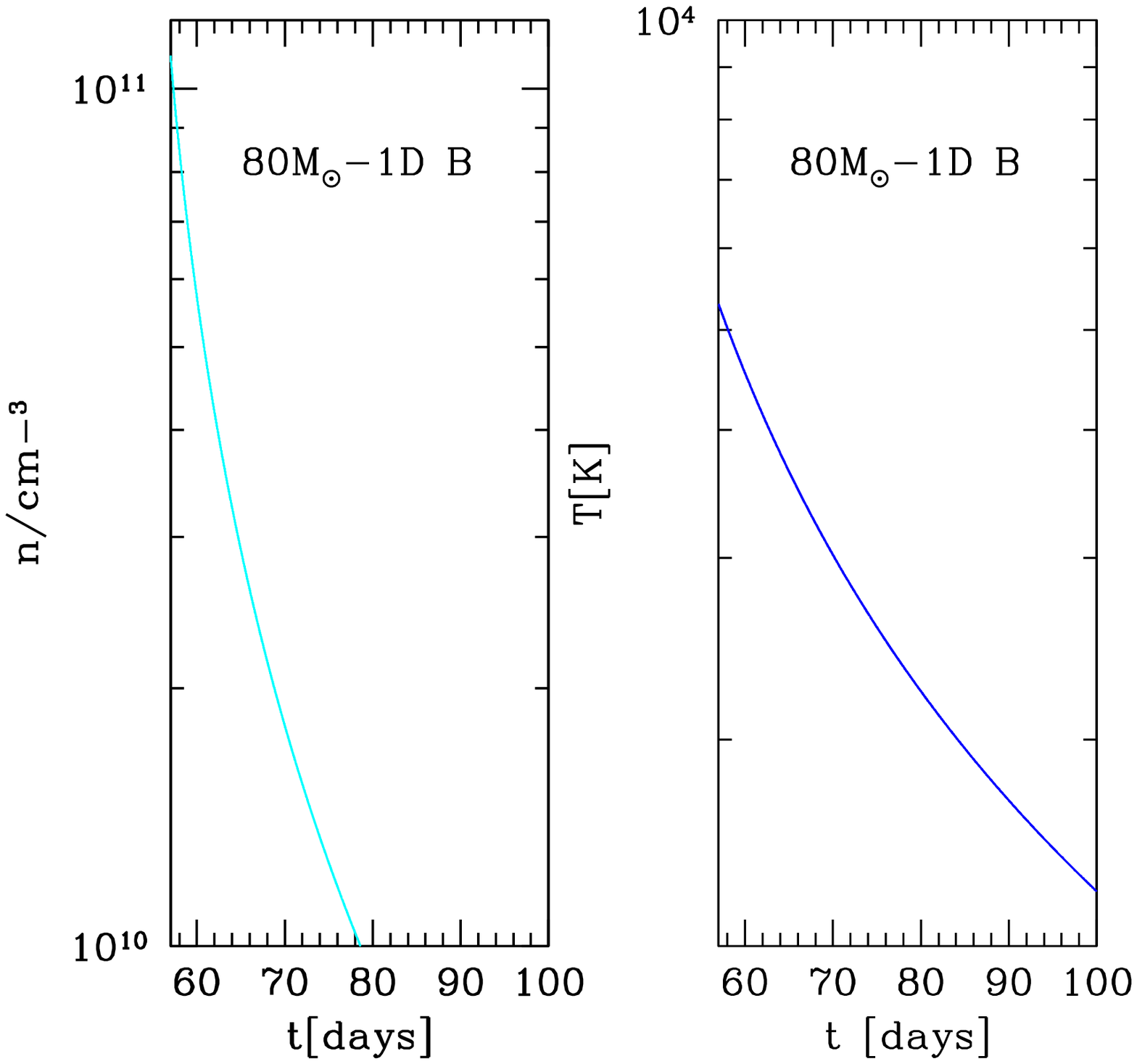}
\includegraphics[angle=0,width=8.5cm]{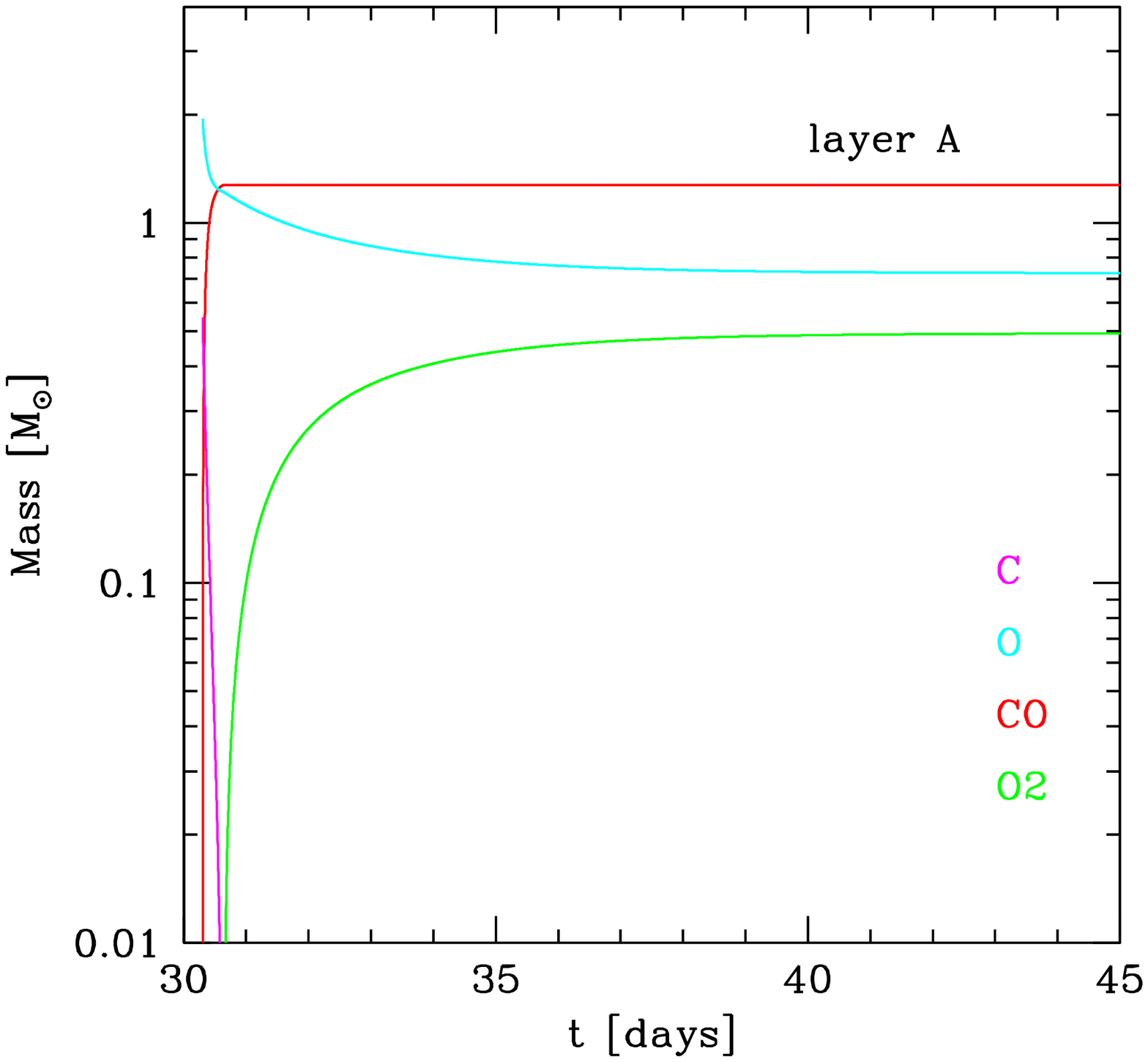}
\includegraphics[angle=0,width=8.5cm]{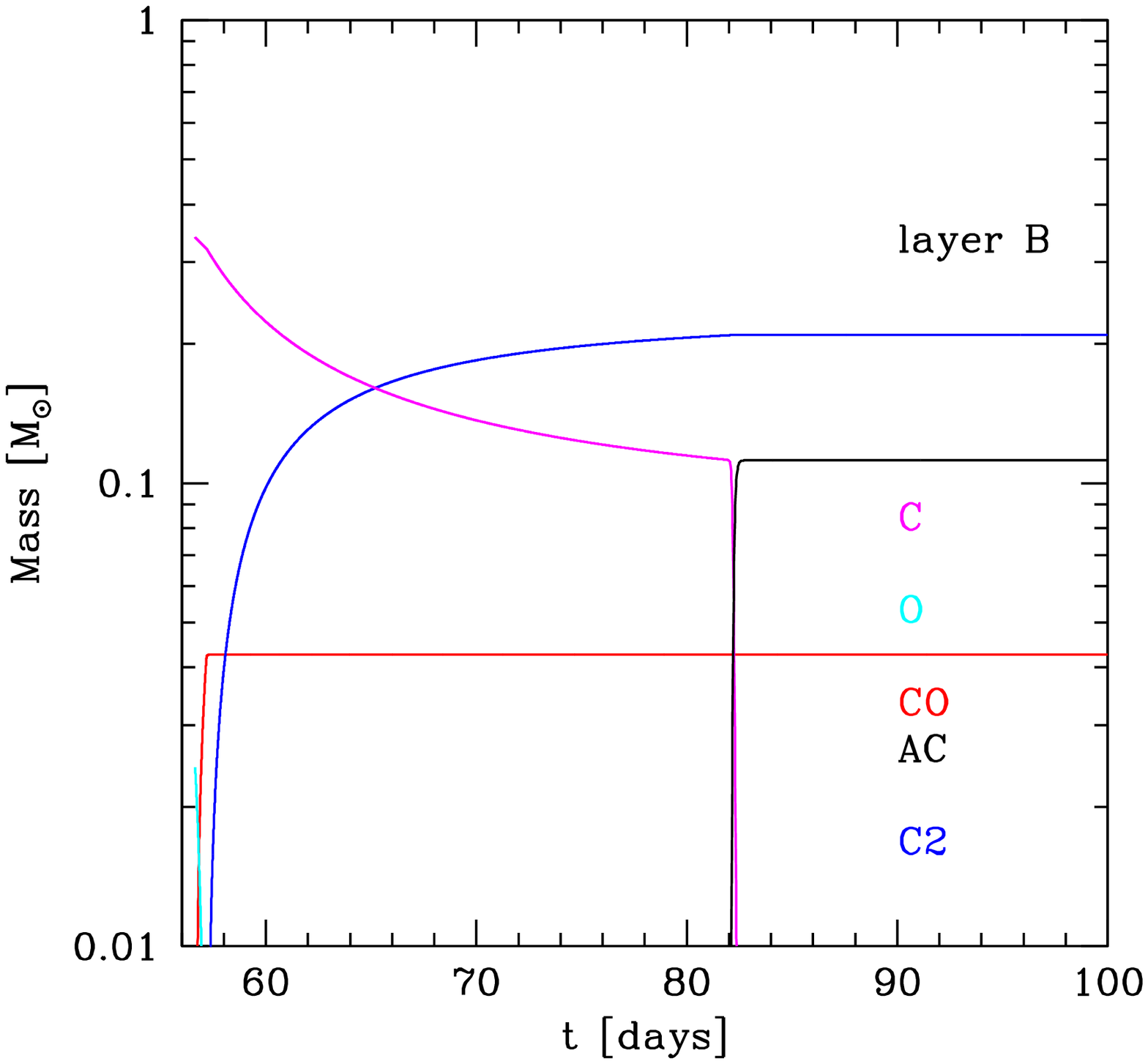}
\caption{Upper panels: temperature and density evolution for 
the stratified $\rm 80 \, M_\odot$-1D (left panels, layer A; right panels, layer B). Bottom
panels: time evolution of the mass in dominant molecular species, carbon 
and oxygen in layer A (left panel) and in  layer B (right panel).}
\label{fig:stratified}
\end{figure*}

Recently, Joggerst et al. (2009) have pointed out that zero 
metallicity supernova progenitors experience more fallback and a smaller degree 
of mixing, induced by Rayleigh-Taylor instabilities, than their solar metallicity 
counterparts. As a result, we expect that the ejecta of faint
SN are likely to be unmixed. Hence, it is important to explore how the mass of dust
estimated above depends on the adopted mixing efficiency. We consider 
a  $\rm 80\, M_\odot$ - 1D SN progenitor model.  The stratified ejecta consists of two layers which 
differ mainly for the different abundance of carbon and oxygen: the inner one (layer A) is characterized 
C/O$ < 1$ whereas in the outer one (layer B) C/O$>1$ (see Table \ref{tab:masses}).
In Fig.~\ref{fig:stratified}, we show the time evolution of the temperature, density, AC, and molecular 
masses formed in the two layers A  and B.  In the inner layer A, 
all the carbon is rapidly locked in CO molecules and consequently C-grains do not form. The oxygen is first depleted in CO molecules
and then in $\rm O_{2}$ molecules. Conversely, in the outer layer B, only $\sim 5\%$ of 
carbon is locked in CO, $\sim 26\%$ of carbon is locked in C$_2$, and the remaining C atoms 
are free to form $\rm 0.112 \, M_\odot$ of AC grains 
as shown in the right, bottom panel in Fig.~\ref{fig:stratified}. $\rm O_{2}$ and $\rm C_{2}$ form mainly 
due to radiative association processes.
Note that in the present formalism, there is no connection between the formation of C2 and the formation
of carbon dust; however, if all the C2 were to transform into carbon dust, the final AC mass would amount 
to $\rm \sim 0.2M_{\odot}$.
In the stratified ejecta, due to the different temperature evolution of the two layers, 
CO molecules start to form at an earlier time than in the corresponding mixed model, 
and the resulting dust mass is $\sim 40 \%$ smaller.

\subsection{Dust destruction by the reverse shock}
\begin{figure*}[t!]
\center{
\plotone{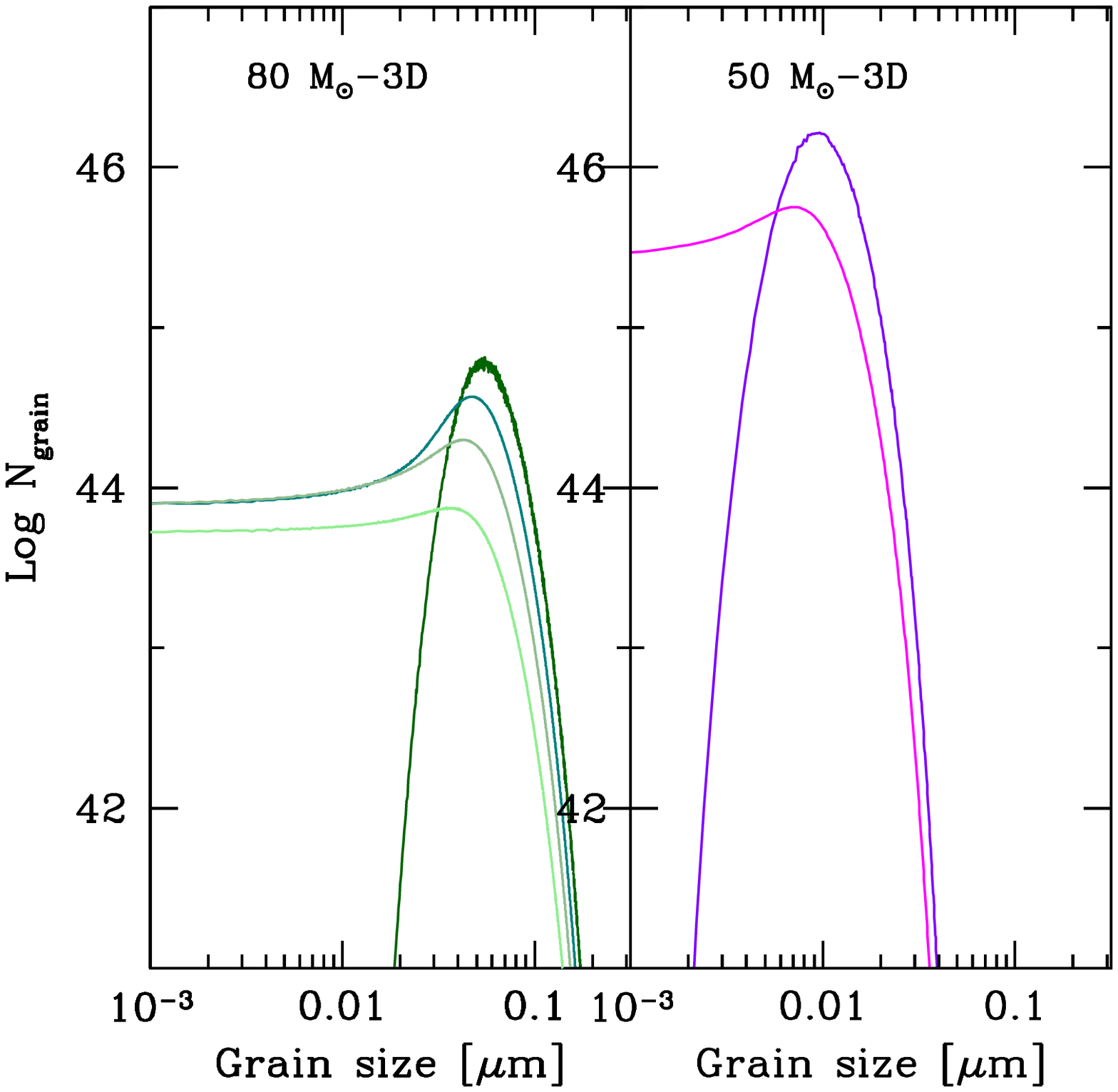}}
\caption{Size distribution for grains formed in the ejecta of 80M$_{\odot}$-3D and 50M$_{\odot}$-3D models, for 
no reverse shock model ({\bf norev}) and for models with a circumstellar medium density of 
$\rm \rho_{ISM} = 10^{-25} \, ({\bf rev1}), 10^{-24}$ \,({\bf rev2}) and $\rm 10^{-23} \, gr\, cm^{-3}$ ({\bf rev3}), from top to bottom.
For the 50 $M_\odot$ model the rev2 and rev3 cases are out of scale.}
\label{fig:graindistri}
\end{figure*}

As the ejecta expand, a forward shock is driven into the interstellar medium, which
compresses and heats the ambient gas. The shocked ambient gas drives a reverse shock in the ejecta, 
which in $\sim 10^3$ years has swept over a considerable fraction of its volume. 
The passage of the reverse shock can be particularly destructive for
the small dust grains, due to the transfer of thermal and kinetic energy during impact with gas particles
(sputtering). We estimate the mass of newly formed dust that is able to survive the passage of the reverse shock
and enrich the interstellar medium following the same model developed by Bianchi \& Schneider (2007). 

The dynamics of the reverse shock is described using the analytical approximations of Truelove \& McKee (1999)
for a uniform density distribution inside the ejecta, that allow to compute the velocity of the shock as a function of the 
SN explosion energy, $\rm E_{kin}$, the mass of the ejecta, $\rm M_{eje}$, and the density of the interstellar medium,
$\rm \rho_{ISM}$. For each SN model, we compute the dynamics of the reverse shock using the same values 
for $\rm E_{kin}$ and $\rm M_{eje}$ that have been used in the dust formation calculations (see Table 1). In addition,
we study the effect of three different ISM environments, with $\rm \rho_{ISM} = 10^{-25}, 10^{-24}$ and $\rm 10^{-23} g\, cm^{-3}$.

We assume that dust grains are distributed uniformly within the ejecta, and that the size distribution is the same everywhere.
In Fig.~\ref{fig:graindistri} we show the grain size distribution that result from grain condensation and accretion in the (fully mixed) ejecta 
of the 50\,M$_{\odot}$ and 80\,M$_{\odot}$ SN models. Initially, the grain sizes follow a log-normal distribution 
in the range $(10^{-2} - 0.5)\, \mu$m. However, in the shells that have been visited by the reverse shock, dust grains are
bathed in a gas heated to high temperature (of the order of $\rm 10^7-10^8 \rm K$). In addition, the gas is slowed down, decouple
from the grains and transfer thermal and kinetic energy to the grains by means of collisions. Following Bianchi \& Schneider (2007),
we consider both thermal and non-thermal sputtering. Note that since the reverse shock velocity, which is of the order of 
$10^3 \rm km/s$, is larger when the SN explodes in a denser circumstellar medium (Truelove \& McKee 1999), 
the reverse shock travels faster inside the ejecta and encounters a gas at higher density. This increases the effect of sputtering.
The effects of the reverse shock on the grain size distribution function is also shown in Fig.~\ref{fig:graindistri}: the larger
grains are eroded and this produces a  shift of the distribution to lower grain sizes,
with a tail extending to a few $\rm \AA$. Note that when the number of monomers in each grain becomes less than 2, 
the grain is evaporated returning the metals into the gas phase (see equation 2 in Bianchi \& Schneider 2007).
 
\begin{figure*}[t!]
\center{
\plotone{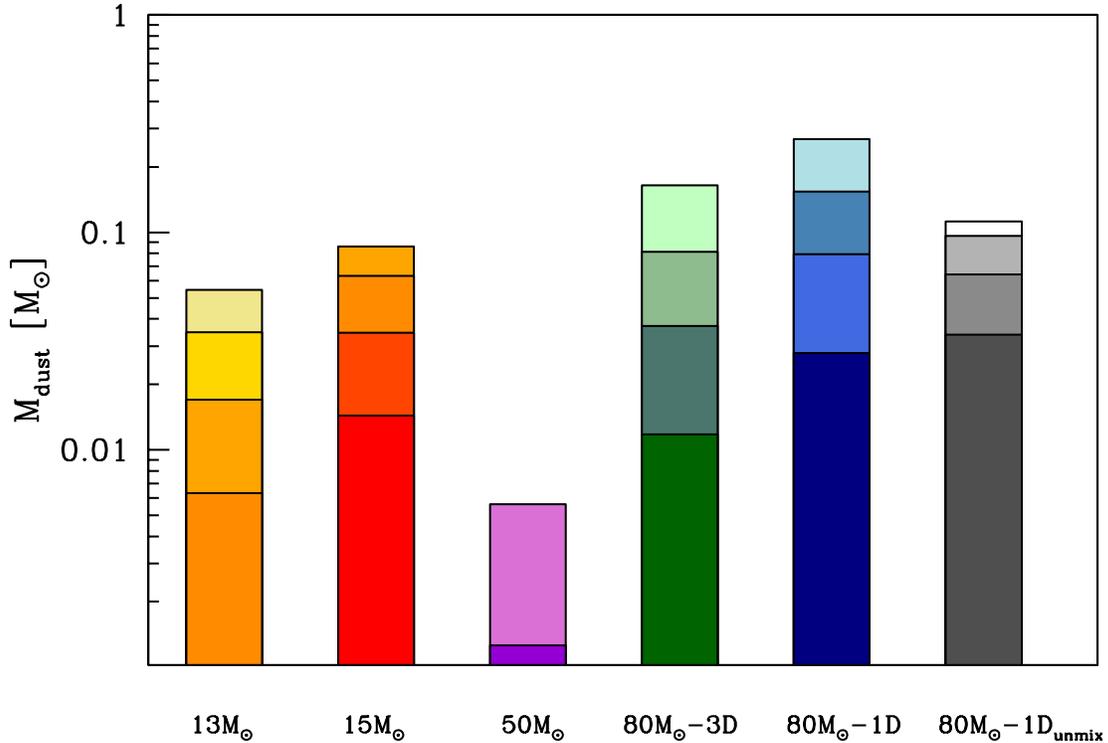}}
\caption{Histograms showing the mass of dust at the end of nucleation, and after the passage of a reverse shock
of increasing intensity for all the faint SN progenitors considered in the present study. 
From top to bottom: no reverse shock models ({\bf norev}), models with a circumstellar medium density of 
$\rm \rho_{ISM} = 10^{-25} \, ({\bf rev1}), 10^{-24}$ \,({\bf rev2}) and $\rm 10^{-23} \, gr\, cm^{-3}$ ({\bf rev3}).
For the 50 $M_\odot$ model the rev2 and rev3 cases are out of scale. }
\label{fig:revshock}
\end{figure*}

The destructive efficiency of the reverse shock is shown in Fig.\ref{fig:revshock}, 
where we plot the mass of dust that survives for increasing shock strengths. 
Depending on the progenitor model, the fraction of newly formed dust that is able to survive ranges between $80 \%$ to a few percent.
These numbers are in agreement with what has been found by other studies using hydrodynamical simulations 
(Nozawa et al. 2007; Silvia, Smith \& Shull 2010, 2012). In particular, by post-processing an idealized numerical simulation
of a planar shock wave impacting a dense, spherical clump, Silvia et al. (2010, 2012) have calculated grain sputtering
for a variety of species and size distributions. By altering the clump overdensity and the shock velocity, they find that 
grains with radii smaller than 0.1 $\mu$m are sputtered to much smaller radii and often completely destroyed and that
the percentage of dust mass that survives critically depends on the initial grain size distribution and on the velocity of
the reverse shock, ranging between complete destruction to $\sim 44\%$ survival for the largest and most robust grains.

Given that the dust produced in faint SN ejecta is composed only by AC grains, the fraction of mass that gets destroyed is returned into the gas phase under the
form of C atoms. The resulting carbon condensation efficiencies, defined as the ratio of the mass of AC grains
relative to the initial mass of C atoms present in faint SN ejecta are shown in Table\,\ref{tab:crit}.
These, together with the gas-phase abundances of the other elements present in the ejecta represent the 
initial conditions for the collapse calculations described in the next section.   

\begin{table*}
\caption{Properties of models which we investigate with the critical abundances of carbon,
$\rm [C/H]_{cr}$, for low-mass star formation.
We here show our models with metal abundances fitted for 3D stellar atmospheric models
by Keller et al.(2014).
$r_{\rm C}^{\rm cool}$ and $f_{\rm C,cond}^{*}$ are, respectively, the characteristic radii of carbon grains
(see text) and the carbon condensation efficiency at density $n_{\rm H} = 10^{12} \ {\rm cm^{-3}}$,
resulting from our collapse calculations with the effects of grain growth.
Carbon abundance is set the observed value, $\rm [C/H]=-2.60$. The values in bold face are below 
the corresponding critical limits, $\rm D_{trans} > -3.5 \pm 0.2$ and  ${\cal D}_{\rm cr} = [2.6 - 6.3] \times 10^{-9}$.}
\label{tab:crit}
\begin{tabular}{@{}cccccccc}
\hline
 SN progenitor & Model & $f_{\rm C,cond}^{\rm ini}$ & ${\cal D}_{\rm ini}$  & $\rm D_{trans,ini}$ & $r_{\rm C}^{\rm cool}$ $\rm (\mu m)$ & $f_{\rm C,cond}^{*}$ & $\rm [C/H]_{cr}$ \\
\hline
13 $\rm M_\odot$ & norev & 0.84  & $9.10 \times 10^{-6}$   & {\bf -4.07}    &     $0.108$      & $    0.84$ & $ -4.65$ \\
          &  rev1    &   0.53                           & $5.77\times10^{-6}$     &  -3.07   &     $0.104$     & $    0.63$ & $   -4.54$ \\ 
        &  rev2    &   0.26                           & $2.83\times10^{-6}$     &   -2.82  &     $0.119$     & $    0.63$ & $   -4.48$ \\
    &  rev3    &   0.10                           & $1.05\times10^{-6}$     &    -2.71 &     $0.148$     & $    0.63$ & $   -4.39$ \\
15 $\rm M_\odot$ & norev   &   0.73                           & $7.90 \times 10^{-6}$   &  {\bf -4.23}   &    $0.177$     & $    0.73$ & $   -4.37$ \\
                 &  rev1    &   0.53                           & $5.78\times10^{-6}$     &  -3.26    &     $0.170$     & $    0.59$ & $   -4.30$ \\
              &  rev2    &   0.29                           & $3.16\times10^{-6}$     &   -2.94   &     $0.191$     & $    0.59$ & $   -4.25$ \\
        &  rev3    &   0.12                           & $1.32\times10^{-6}$     &    -2.80  &    $0.228$     & $    0.59$ & $   -4.17$ \\
50 $\rm M_\odot$       & norev   &   0.0056                       & $6.10 \times10^{-8}$    &   -3.41   &      $0.014$     & $  0.0059$ & $   -3.38$ \\
            &  rev1   &   0.0013  & $1.38\times10^{-8}$      &  -3.40    &     $0.012$    & $  0.0013$ & $   -2.79$ \\
           &  rev2   &   0.0004  & ${\bf 4.26\times10^{-9}}$      &   -3.40   &      $0.012$    & $  0.0004$ & $   {\bf -2.30}$ \\
          &  rev3   &   0.0001   &${\bf 1.30\times10^{-9}}$      &   -3.39    &    $0.012$    & $  0.0001$ & $   {\bf -1.77}$ \\
80 $\rm M_\odot$-3D  & norev  &   0.15                            & $1.64 \times10^{-6}$    &    -3.25   &    $0.073$    & $    0.15$ & $   -4.07$ \\
            &  rev1   &   0.08                          & $8.07\times10^{-7}$     &    -3.12    &     $0.067$    & $    0.08$ & $   -3.80$ \\
     &  rev2   &   0.03                          & $3.69 \times10^{-7}$    &     -3.07    &    $0.064$    & $    0.03$ & $   -3.49$ \\
              &  rev3   &   0.01                          & $1.17\times10^{-7}$     & -3.04 &    $0.061$     & $    0.01$ & $   -3.01$ \\
\hline
\end{tabular}
\end{table*}

\section{The star forming environment of SMSS J031300}

In this section, we test the hypothesis that SMSS J031300 has formed 
out of the collapse and fragmentation of a gas cloud that has been pre-enriched
by the ejecta of one of the Pop III faint SN studied in the previous section.
We first discuss the expansion of the ejecta into the surrouding ISM and then
the subsequent recollapse of the star forming cloud, to assess the relative role of 
fine-structure-line and dust cooling for the formation of SMSS J031300. \\

\subsection{Ejecta expansion and mixing with the ISM} 
The mixing and fallback model - as first introduced  
by Umeda \& Nomoto (2002) - naturally predicts a
stratified ejecta: the inner material is assumed to be 
mixed by the growth of Rayleigh-Taylor instabilities 
during shock propagation, while the ejecta 
is dominated by the external material 
plus a small fraction of the mixed gas near the mass-cut. 
The mixing efficiency  is still a matter of debate due to the complexity of the 
mixing/fallback scenario, which depends on the explosion energy, 
the gravitational potential, and the asphericity during the collapse.
This is the reason why we have studied dust formation in both perfectly
mixed and stratified ejecta. Yet, when we compare the observed surface
elemental abundance of a stellar fossil, such as SMSS J031300, with
the yields of a single Pop III SN, we make the implicit assumption that
the ejecta material has uniformly enriched the parent star forming cloud.
Hence, whatever is the degree of mixing at
nucleation,  45-80 days after the explosion, or after the passage of
the reverse shock, $(2.6 - 3.4) \times 10^3$ years later, the ejecta
material needs to be perfectly mixed into the star forming cloud
before its subsequent collapse. 

An estimate of the maximum timescale available to fully mix the enriched material 
can be obtained by computing the total mass and radius of the cloud, M$_{\rm cloud}$, 
in which the ejecta needs to be diluted in order to match the observed metallicity 
on the surface of SMSS J031300, $\rm Z_{\rm obs} \sim 2.67 \times 10^{-3}\, Z_\odot$ \citep{Kel2014},
\begin{equation}
\rm M_{\rm cloud} = \frac{\rm M_{\rm met, ejecta}}{Z_{obs} X_H}
\end{equation}
where $\rm X_H = 0.75$ and we have assumed an ISM with primordial composition.
We find that M$_{\rm cloud}$ is an increasing function of the progenitor mass, 
since more massive progenitors have larger metal yields, and ranges between 
$10^4$ and $2 \times 10^5\, \rm M_\odot$. This gas mass is roughly consistent 
with the baryonic content of a $\rm \sim 10^6 M_\odot$ mini-halo.
The radius of the cloud depends on the density of the ISM ($\rm \rho_{ISM}$): 
to be consistent with the reverse shock models described in the previous section, we 
assume values in the range $\rm [10^{-26} - 10^{-23}] \,gr\, cm^{-3}$, where the lower limit 
approximately corresponds to the critical density at $z = 10$. The corresponding cloud
radius is, 
\begin{equation}
\rm R_{\rm cloud} = \left[\frac{3\rm M_{\rm cloud}}{\rm 4\pi \rho_{\rm ISM}}\right]^{1/3}
\end{equation}
and varies in the range  $\rm 20\,pc \le  R_{cloud} \le 100\,pc$, which is reached by the SN shock in its
Sedov-Taylor phase on a timescale, $\rm t_{exp} = R_{cloud}^{5/2} \, (\rho_{ISM}/E_{exp})^{1/2} \sim 0.2 - 10$~Myr. 

Following Madau, Ferrara \& Rees (2001), we can estimate the growth timescale of Rayleigh-Taylor instabilities 
as $\rm t_{RT} \approx (l_s /2 \pi g)^{1/2}$, where $\rm g$ is the gravitational acceleration and $l_s$ 
is the spatial scale to be mixed. Using the values computed above for the cloud mass and radius, we find
$\rm t_{RT} = 0.4 - 3$~Myr, depending on the ISM density. 
Hence, in less than a few Myr after the explosion of a faint Pop III SN, the ejecta material may have already 
mixed and enriched the gas out of which second generation stars form. 
The resulting stellar mass depends on the cooling and fragmentation properties of the collapsing gas, which 
we investigate in the following section.\\

\subsection{Cooling and fragmentation of the collapsing cloud}
We perform a one-zone semi-analytic collapse calculation
to follow the temperature evolution of the cloud \citep{Omu2000,Omu2005},
setting the abundances of heavy elements and dust species as inferred from the SN dust models described
in the previous section. We refer the interested reader to Chiaki et al. (2014) for a detailed 
description of the model.

In this collapse calculation, we solve non-equilibrium chemistry of 27 gas-phase species
(H$^+$, $e^-$, H, H$^-$, H$_2$, D$^+$, D, HD,
C$^+$, C, CH, CH$_2$, CO$^+$, CO, CO$_2$, O$^+$, O, OH$^+$, OH, 
H$_2$O$^+$, H$_2$O, H$_3$O$^+$, O$_2^+$, O$_2$, Si, SiO, and SiO$_2$)
with a chemical network of 55 reactions. Radiative cooling owing to atoms/ions (C{\sc i}, C{\sc ii}, and O{\sc i}) and molecules
(H$_2$, HD, CO, OH, and H$_2$O), and the heating/cooling by H$_2$ formation/destruction are computed
self-consistently at each time. Gas cooling by heat transfer between gas and dust is calculated as the sum over
all dust species and size bins. In addition to the dust mass that is present in the gas cloud at the onset of collapse,
we also treat accretion of gas-phase metal species onto dust grains during the collapse (grain growth, Nozawa et al. 2012; Chiaki et al. 2014).
For the purpose of the present analysis, we set the cosmic microwave background (CMB) temperature at
$\rm T_{CMB} = 27.3~K$, corresponding to $(1+z) = 10$ (see Schneider \& Omukai 2010 for the effects of the CMB temperature
floor on the collapse of clouds at different metallicities).

\begin{figure*}[t!]
\hspace{-1.0cm}
\includegraphics[angle=0,width=7.25cm]{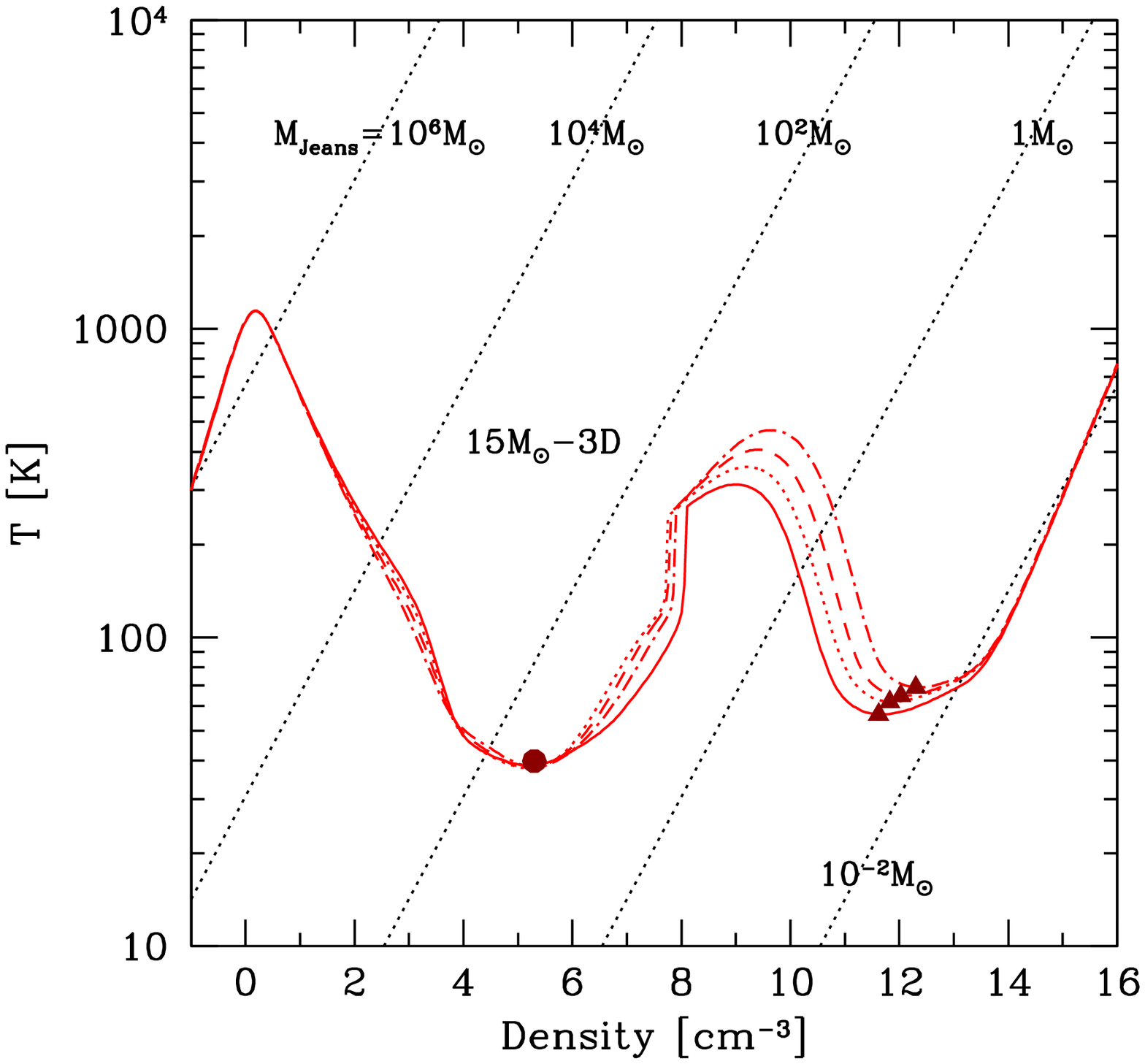}
\hspace{-2.35cm}
\includegraphics[angle=0,width=7.25cm]{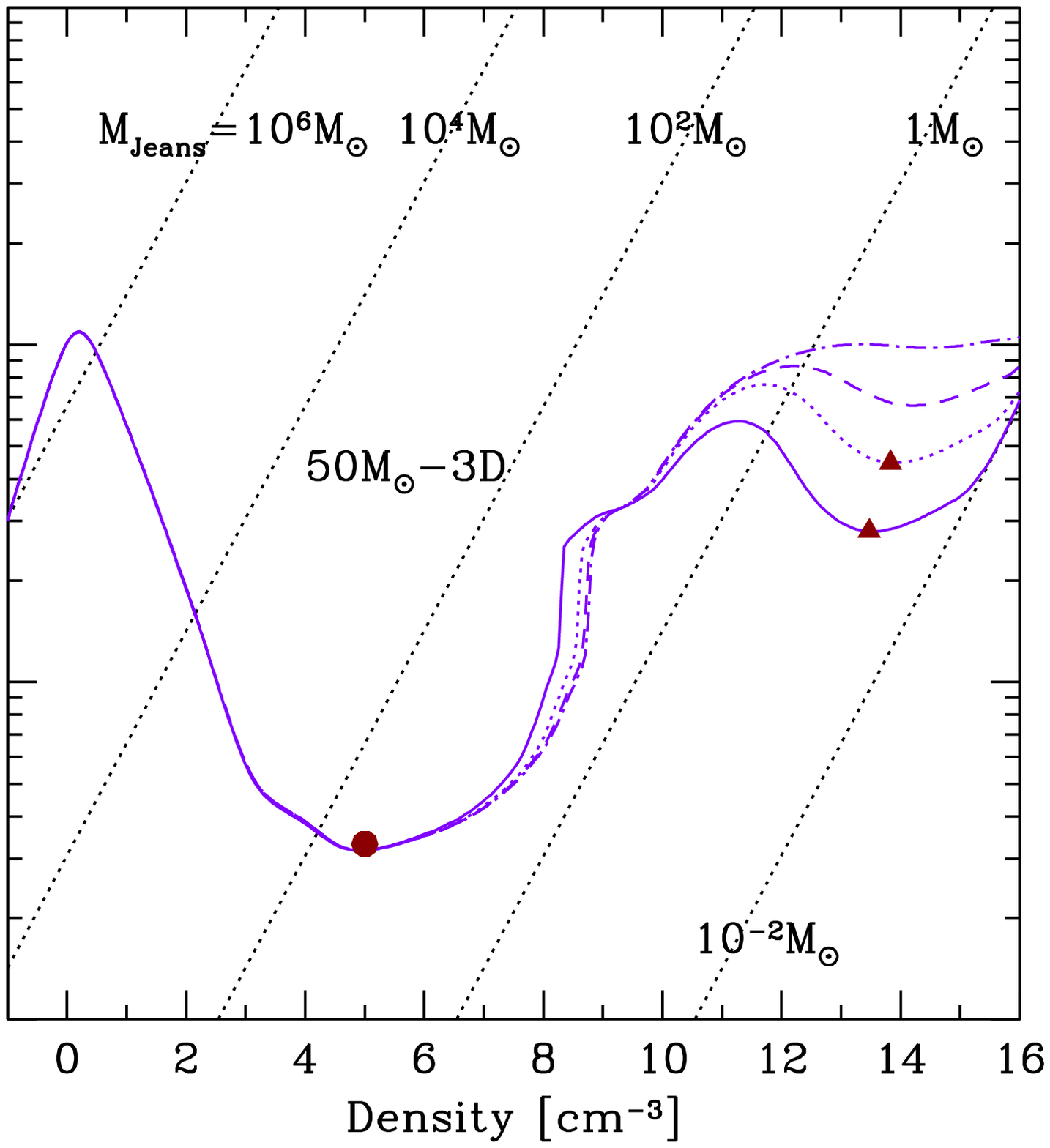}
\hspace{-2.35cm}
\includegraphics[angle=0,width=7.25cm]{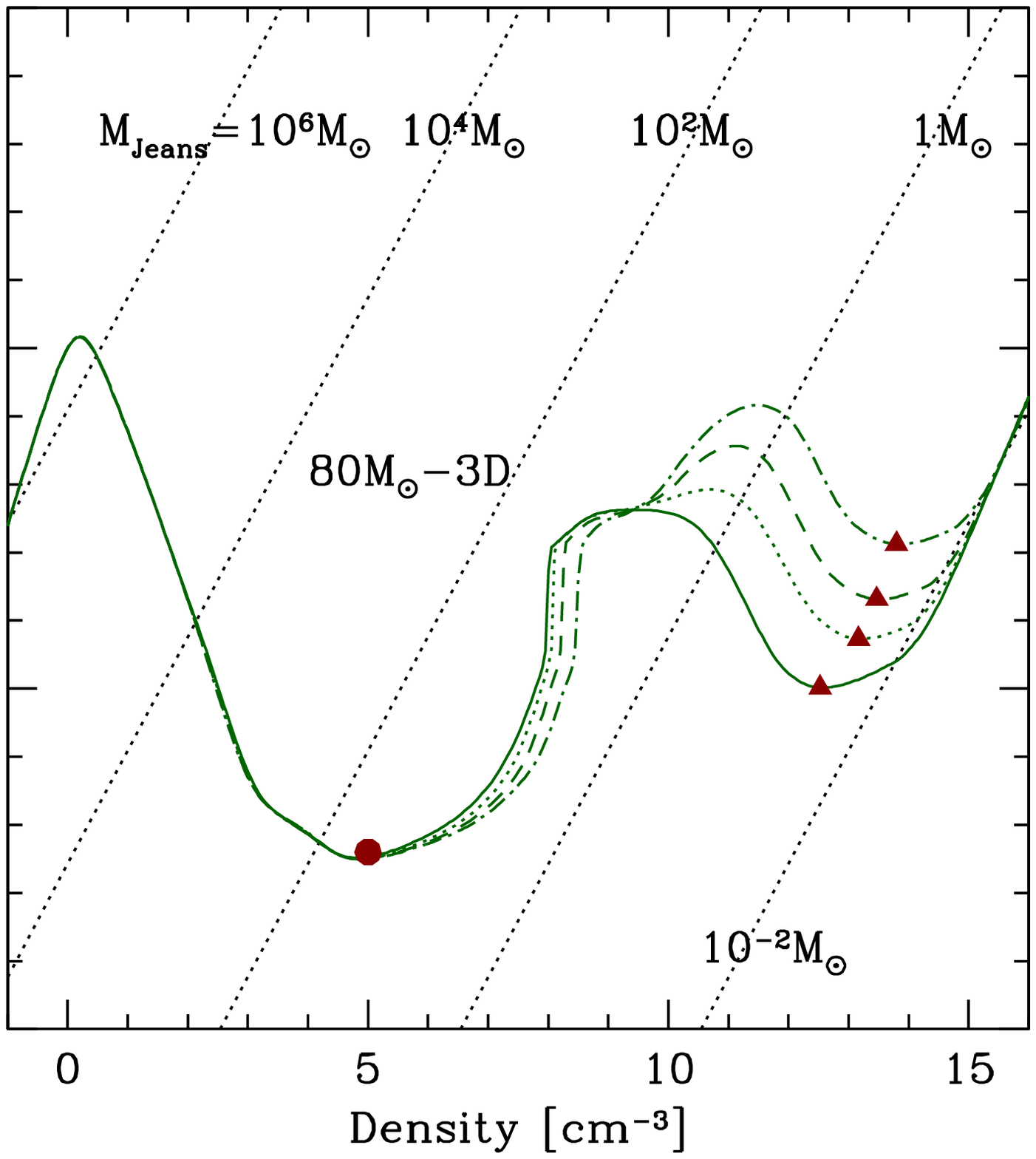}
\caption{Temperature evolution as a function of the central density of clouds 
pre-enriched by Pop III faint SN explosions of 
$\rm 15 \ M_\odot$ (left), $\rm 50 \ M_\odot$ (middle), and $\rm 80 \ M_\odot$-3D (right) stars.
Solid curves represent  models with no reverse shock, dotted-, dashed- and dot-dashed curves 
represent models with a circumstellar medium density of 
$\rm \rho_{ISM} = 10^{-25}, 10^{-24}$ and $\rm 10^{-23} gr \, cm^{-3}$, hence an increasing
reverse shock destructive efficiency. Red triangles mark at the states where gas fragmentation 
conditions are met.}
\label{fig:nT}
\end{figure*}

Fig.~\ref{fig:nT} shows the cloud evolution in the temperature-density plane. Each panel 
represents the results assuming that the cloud has been pre-enriched by the explosion of 
an individual Pop III faint SN model. We have selected the results of three different 
progenitor masses (the 13 and 15 $\rm M_{\odot}$ models lead to very similar results)  
adopting dust yields formed in uniformly mixed ejecta (3D models, although 
the results are not affected by the set of reference abundances adopted in the explosion calibration). 
For each of these, we consider the decreasing dust-to-metal ratios that emerge from
the passage of the reverse shock of increasing intensity. The initial dust-to-gas mass ratio at the
onset of collapse, ${\cal D}_{\rm ini}$ and initial transition discriminant $\rm D_{\rm trans,ini}$ 
are shown in Table \ref{tab:crit}, where we have written in bold face only the values
that are below the corresponding critical limits. Hence, we would expect metal-line cooling
to be effective for all but models 13 and 15 $\rm M_{\odot}$ in the no reverse shock case, 
and dust cooling to be effective for all but models $\rm 50 \, M_{\odot}$ in the two stronger 
reverse shock cases (rev2 and rev3).  

We can compare these simple expectations with full collapse calculations.
The circles and triangles in the figure mark the states where fragmentation
conditions are met \citep{Sch2010}. At densities 
$n_{\rm H} \sim 10^{5} \ {\rm cm^{-3}}$, fragmentation is due to
line-cooling. In the 13 and 15\,$\rm M_{\odot}$ models, the dominant coolants are HD and C{\sc i}.
For models with moderate reverse shock destruction (norev, rev1),
$f_{\rm C,cond}^{\rm ini} >0.5$ and - as a result - 
H$_2$ formation on grain surface is very efficient, enabling the formation of HD molecules.
On the other hand, C{\sc i} cooling efficiency is comparable 
to (3D cases) or exceeds (1D cases, where [C/H] is larger than 3D) HD 
cooling for rev2 and rev3 models\footnote{In our calculations, gas cooling is dominated by C{\sc i} 
because we are neglecting the presence of a Far-UV (FUV) radiation
field. If some FUV is present, C{\sc i} is photo-ionized and - as a result - C{\sc ii} dominates
cooling. However, this does not affect our conclusions as the two cooling rates are similar.}.
For the larger progenitor mass models (50 and $\rm 80 \ M_\odot$), the dominant
coolant at intermediate densities is C{\sc i}  as a result of the lower initial condensation efficiencies.
Yet, as shown by the diagonal lines in the three panels, the characteristic masses of the fragments
formed in this early stage of  collapse (the Jeans mass at the density where instability 
occurs) are estimated to be $\rm \sim 100 \ M_\odot$.

It is important to note that
although sputtering of AC grains release C atoms in the gas phase that may be
potentially accreted onto the surviving grains at a later stage of the collapse, 
grain growth has a non-negligible impact on the results only if $\rm C/O > 1$ at the onset
of collapse. In fact, AC grain growth is prevented by efficient formation of CO 
molecules at $n_{\rm H} \sim 10^{4} \ {\rm cm^{-3}}$ \citep{Chi2014}. 
This is confirmed by comparing the carbon condensation efficiencies at the onset
of collapse, $f_{\rm C,cond}^{\rm ini}$, and a later stage of the collapse,  $f_{\rm C,cond}^{*}$,
when the density is large enough for grain growth to occur efficiently ($n_{\rm H } = 10^{12} \ {\rm cm^{-3}}$,
see below). The values predicted by each model are shown in Table \ref{tab:crit}; 
as expected, grain growth has non negligible effect only for the 13 and 15 $\rm M_\odot$ models
with reverse shock. 
Hence, for all but the smallest progenitor mass models,
the conditions for dust cooling and fragmentation are largely
determined by the dust-to-gas ratio initially present in the cloud as a result of the SN explosion
and the associated reverse shock. 

Fig.~\ref{fig:nT} shows that dust cooling is efficient at high densities
$n_{\rm H} = 10^{10}$--$10^{14} \ {\rm cm^{-3}}$  triggering fragmentation and 
potentially leading to the formation of a low-mass star as SMSS J031300.
The only two unsuccessful cases are associated to the $\rm 50 \ M_\odot$ faint SN
explosion with the stronger reverse shock models. Not surprisingly, in these two
models the amount of dust that survives the passage of the reverse shock is not
enough to keep the initial dust-to-gas mass ratio above the critical one, 
${\cal D}_{\rm cr} = [2.6 - 6.3] \times 10^{-9}$
(see Table \ref{tab:crit}). \\

Finally, it can be shown that the full collapse calculations confirm that even in
unmixed models, which are characterized by smaller dust yields, dust-driven
fragmentation conditions are met for all reverse shock cases considered, consistent
with the fact that the initial dust-gas ratios are super-critical, varying 
in the range ${\cal D} =  7.94 \times 10^{-8} - 1.11 \times 10^{-6}$. 

Following Chiaki et al. (2014), we can estimate the critical elemental
abundances that enable dust-induced fragmentation. 
For CEMP stars enriched by faint Pop III SN explosions, 
carbon grains are the dominant species in the parent cloud; therefore, 
the critical dust-to-gas ratio can be easily associated to a critical 
carbon abundance, 
\[
A_{\rm C,cr} = 1.4 \times 10^{-3} \frac{4 \, s_{\rm C} \, r_{\rm C}^{{\rm cool}}}{3X f_{\rm C,cond}(t) \, \mu _{\rm C}}
\] \[
~~~~~~~~~\times 
\left( \frac{T}{10^3 \ {\rm K}} \right) ^{-1/2}
\left( \frac{n_{\rm H}}{10^{12} \ {\rm cm^{-3}}} \right)^{-1/2},
\]
where $s_{\rm C}$ is the material density of carbon grains ($s_{\rm C} = 2.28 \ {\rm gr \ cm^{-3}}$)
and $r_{\rm C}^{\rm cool} = \langle r^3 \rangle _{\rm C} / \langle r^2 \rangle _{\rm C}$
is the characteristic grain radius for gas cooling \citep{Chi2014}.
Table \ref{tab:crit} shows the cooling radii and the critical carbon abundances normalized to the solar value
($A_{\rm C , \odot}=2.69 \times 10^{-4}$ taken from the photospheric value in Asplund 2009).
Since the observed carbon abundance of  SMSS J031300 is $\rm [C/H] = -2.60$ \citep{Kel2014}.
the critical conditions are satisfied by all the models but the $\rm 50 \ M_\odot$ rev2 and rev3. 
This is consistent with the full collapse calculations. Hence we conclude that the  formation of the CEMP star
SMSS J031300, even with the lowest iron-content currently observed, may have been triggered by dust cooling and
fragmentation.

\section{Discussion and Conclusions}

The results presented in the previous section have been obtained using 
the dust and metal yields predicted for faint Pop III SN explosions tailored 
to reproduce the observed elemental abundances of SMSS J031300.
For the first time, we have studied the process of dust formation in 
carbon-rich, iron-poor ejecta that characterize these explosions. We
find that carbon can be significantly depleted onto AC grains, with 
condensation efficiencies which can be as large as 75\%. Due to the
small abundance of other refractory elements in the ejecta, silicates,
alumina or magnetite grains do not form in the ejecta of faint Pop III
SN. For most of the models that we have explored, when normalized to the observed
[C/H] which characterize CEMP stars, the predicted dust-to-gas ratios
in the parent star-forming cloud are large enough to trigger dust cooling
and fragmentation, even without the contribution of grain growth. 

If these results can be considered as representative for the general class
of CEMP-no stars, which represent 90\% of CEMP with $\rm  [Fe/H] < -3$, then
our study suggests that dust cooling and fragmentation may have been
responsible for the formation of these low-mass stars. 

In a recent work, Ji et al. (2014) have suggested that the early formation
of low-mass stars may follow two alternative pathways: in the dynamic pathway, 
many sub-clumps are formed through fine-structure line cooling. 
Dynamical interactions can lead to the ejection of a protostar from the parent 
cloud, so that a low-mass star can form even in the absence of dust cooling. 
In the thermal pathway, fragmentation due to fine-structure cooling is inhibited by 
the small abundance of gas-phase elements, and  the entire cloud collapses.
A high-density protostellar disk is formed at the center of the cloud, which then fragments and forms
low-mass stars due to dust cooling. Observational evidence for these two alternative 
pathways comes from the distribution of a sample of metal-poor halo stars and 
dwarf galaxy stars with $\rm [Fe/H] < -3.5$ in a plane where on the y-axis
is the transition discriminant for line-cooling, $\rm D_{trans}$ and on the
x-axis is the [Si/H] (see their Fig. 6); the latter quantity is used as an indication of the
dust-to-gas ratio in the parent cloud, assuming that the dust is mostly
made of silicates. When compared with the critical values for fine-structure-line 
and dust cooling, they find that stars that have a [Si/H] too low to
form through silicon-based dust cooling are characterized by a large transition
discriminant, and satisfy the criterium for fine-structure line cooling and viceversa.

Our study questions this interpretation as it shows that CEMP stars form in C-rich
gas, where  silicates do not form and carbon dust is the dominant dust species.
Hence, for CEMP stars, [C/H] is a better proxy for dust cooling than [Si/H].
In addition, we also show that fine-structure-line and dust-cooling are not
mutually exclusive: the collapsing gas cloud undergoes two distinct phases
of fragmentation, one driven by line cooling (either by metal lines, such as C{\sc i}, C{\sc ii}
or by molecular transitions of H$_2$ and HD) at 
$n_{\rm H} \sim 10^{4} \ {\rm cm^{-3}}$ which leads to the formation of $\sim \rm 100\, M_{\odot}$
sub-clumps; and the other, driven by dust cooling  at $n_{\rm H} = 10^{10}$--$10^{14} \ {\rm cm^{-3}}$,
which triggers the formation of low-mass stars (see also Tanaka \& Omukai 2014).
Thus, the thermal pathway may be responsible for the formation of both C-rich and C-normal
stars, even at lowest [Fe/H] currently observed, although the dynamic pathway may operate as well.

\begin{figure*}[t!]
\hspace{-1.0cm}
\includegraphics[angle=0,scale=0.5]{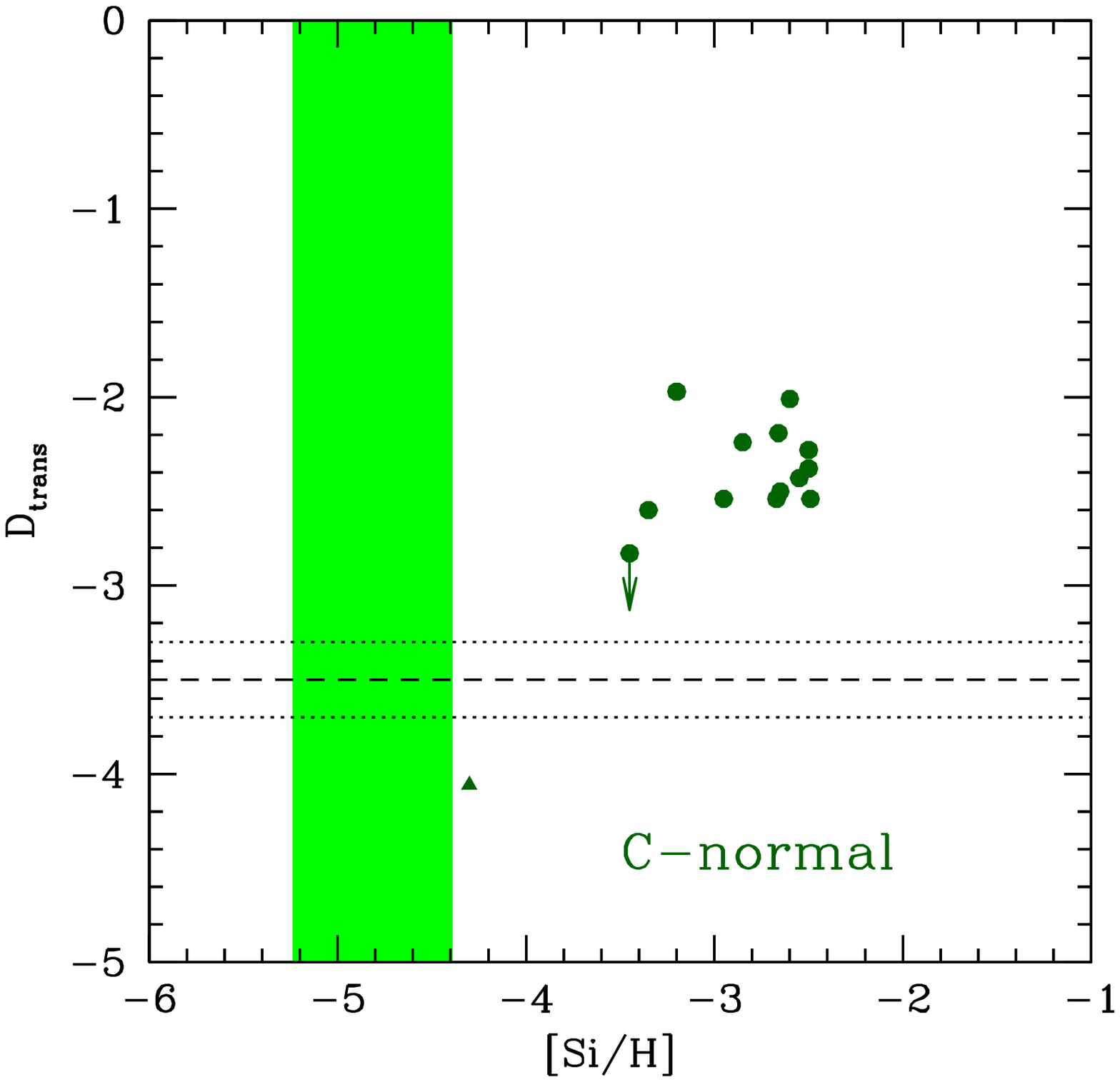}
\hspace{-1.5cm}
\includegraphics[angle=0,scale=0.5]{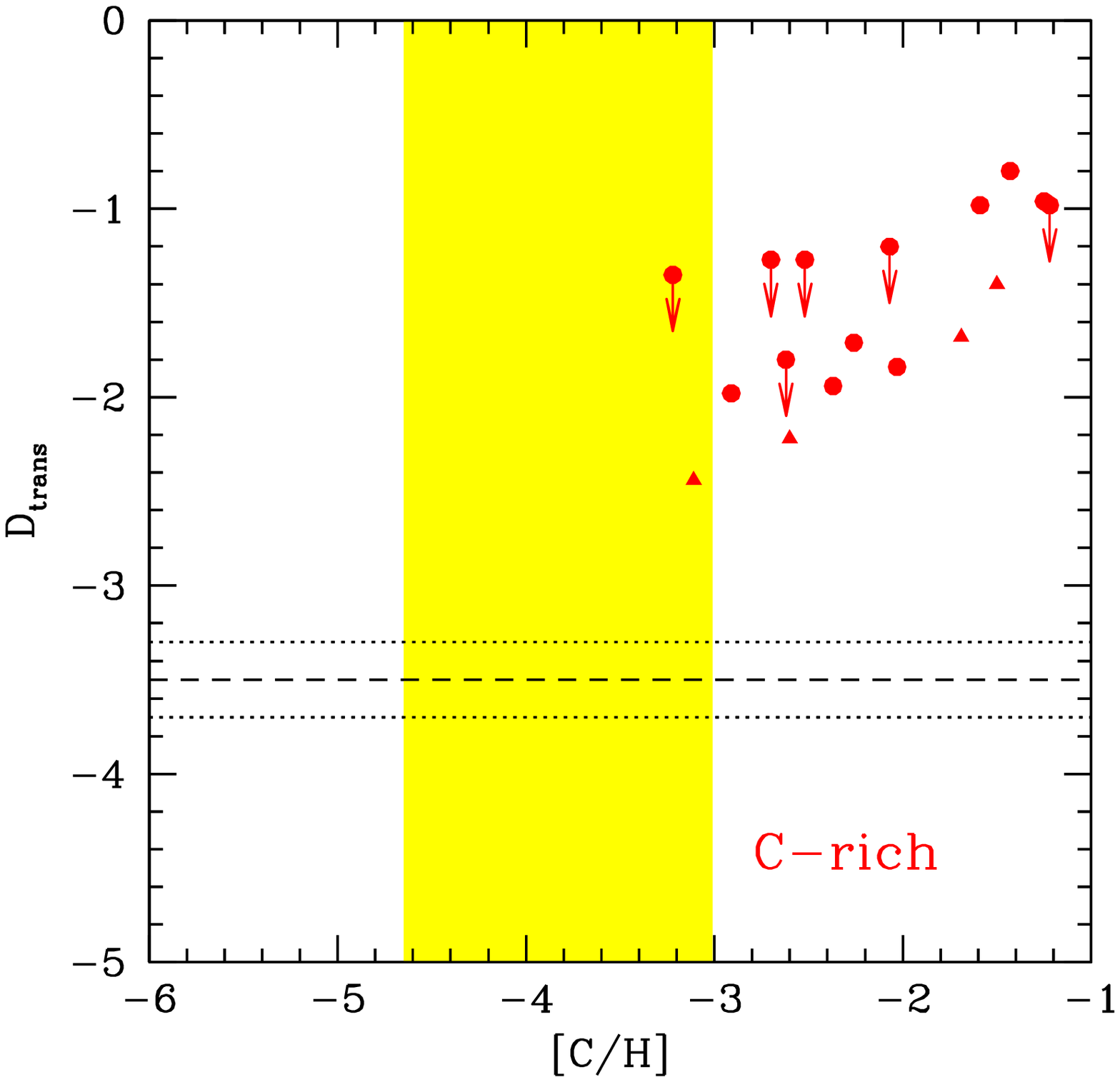}
\caption{Conditions for the formation of low-mass C-normal (left panel) and C-rich stars (right panel).
Observations are taken from $\rm [Fe/H] < -3 $ sample of Norris et al. (2013). Points with triangles
indicate the sub-sample of the 5 ultra-iron poor stars, with $\rm [Fe/H] < -4$ currently known. 
The horizontal lines indicate the critical transition discriminant $\rm D_{trans, cr} = -3.5 \pm 0.2$
for fine-structure line cooling (Frebel \& Norris 2013). 
The vertical shaded regions indicate the range of critical Si (left panel) and C (right panel) 
abundances that would enable Si-dominated dust and C-dominated dust
to activate dust cooling (see text). The arrows indicate stars where either the C or the O
abundances was not measured and only an upper limit was available.}
\label{fig:CSicr}
\end{figure*}

\begin{figure*}[t!]
\center{
\includegraphics[angle=0,scale=0.8]{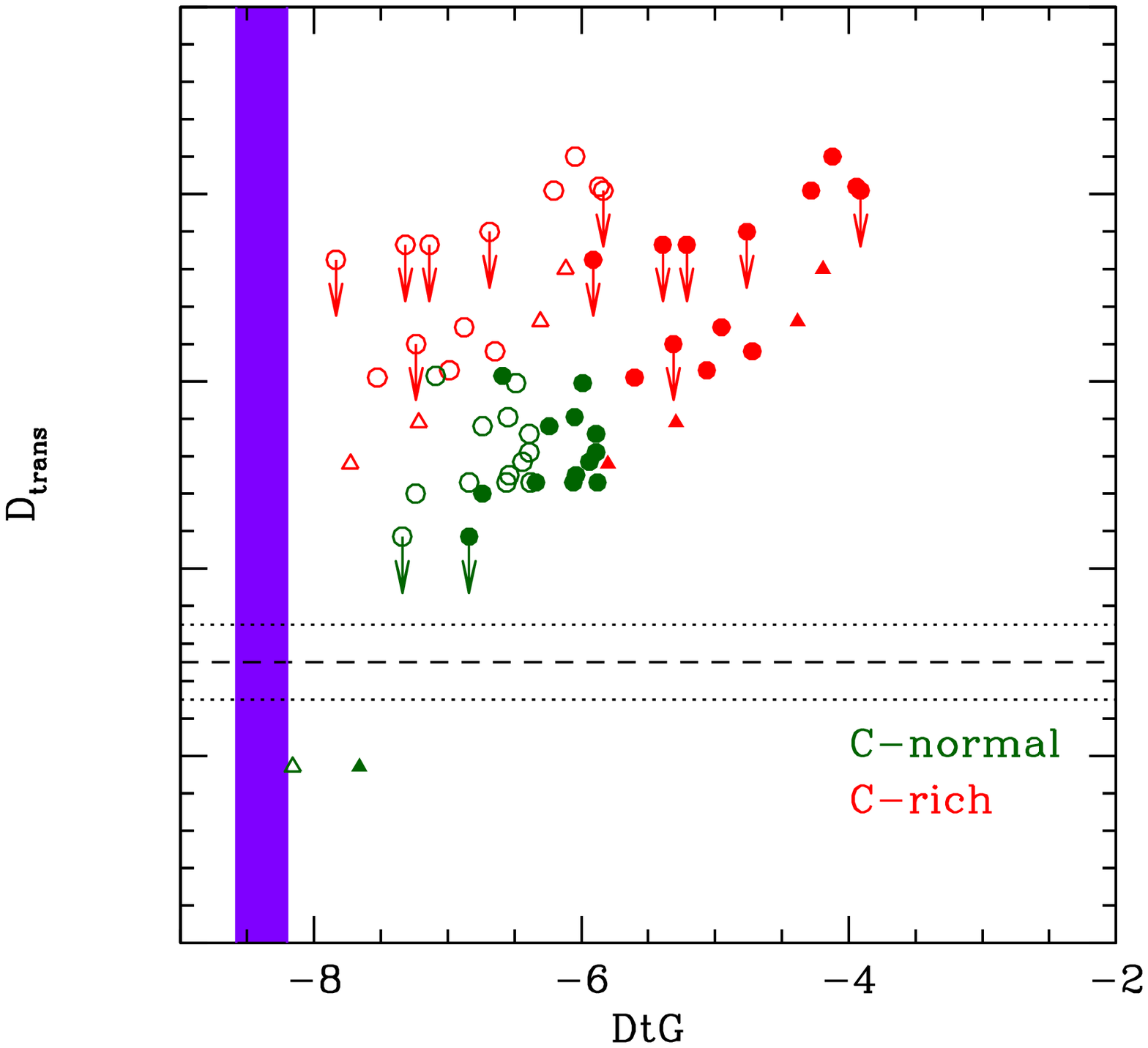}}
\caption{Same as in Fig.~6 but the data are not plotted as a function of the
estimated dust-to-gas mass ratio; for each star we show two values of the dust-to-gas mass
ratio, obtained using the maximum (filled data points) and minimum (empty data points) condensation
efficiencies for C (C-rich) and Si (C-normal) stars (see text). The vertical shaded region represent
the critical dust-to-gas ratio of ${\cal D}_{\rm cr} =  [2.6 - 6.3]\times 10^{-9}$ for dust-cooling (Schneider et al. 2012a). }
\label{fig:dcr}
\end{figure*}

Unfortunately, it is not straightforward to estimate the dust-to-gas ratio of the parent gas cloud
from the observed elemental abundances, similar to what is done for the transition discriminant.
In fact, the dust properties (its composition and size distribution) depend on the nature of the
SN explosions that have dominated the early enrichment and on the level of destruction suffered
by the newly formed dust by the passage of the reverse shock, although the latter process may
be mitigated by grain growth during the collapse. Using the results that we have obtained in
the present study as a proxy for the general class of C-rich metal-poor stars and the results obtained by
Schneider et al.~(2012b) and Chiaki et al.~(2014) for the stars SDSS J1029151 as a proxy for the
general class of C-normal metal-poor stars, we can draw the following guidelines: for C-rich stars,
carbon condensation efficiencies into dust of $0.01 - 0.84$ translate into critical carbon
abundances of  $\rm -4.65 < [C/H]_{cr} < -3.01$; for C-normal stars, the dominant
dust compounds are forsterite (Mg$_2$SiO$_4$) and enstatite (MgSiO$_3$), with 
Mg and Si condensation efficiencies which range between $0.2 - 0.6$ and
associated critical abundances of $\rm -5.02 < [Mg/H]_{cr} < -4.73$ and $\rm -5.24 < [Si/H]_{cr} < -4.73$
(see Table 1 in \citealp{Chi2014}). 

In Figs.~6 and 7, we compare these predictions with observational data taken from the sample 
of C-normal and C-rich stars with $\rm [Fe/H] < -3$ investigated by Norris et al. (2013). 
As shown in Caffau et al. (2011), SDSS J1029151 is the only star, discovered until now, that falls 
in the forbidden zone for fine-structure line cooling (indicated by the triangle below ${\cal D}_{\rm cr}$ 
in Fig.\ref{fig:CSicr}, \ref{fig:dcr}) and no stars appear to fall in the forbidden zone for dust-cooling 
although there are two C-rich stars, HE0557-4840 (Norris et al. 2007) 
and  HE0057-5959 (Yong et al. 2012), which appear to be close to the critical range.

If we represent the critical conditions in terms of the
dust-to-gas ratio, as in Fig.~\ref{fig:dcr}, we see that C-normal and C-rich stars do not appear to follow
different distributions on the plane, although - as expected - the C-normal stars have systematically
lower $\rm D_{trans}$ and lower dust-to-gas-ratios, ${\cal D}$. 

More importantly, the two criteria for low-mass star formation do not appear to be mutually 
exclusive, but rather point to a single pathway for the formation of the first low-mass stars 
in the Universe. 

\acknowledgments
We thank Simone Bianchi for his kind collaboration.
The research leading to these results has received funding from the European Research Council 
under the European Union’s Seventh Framework Programme (FP/2007-2013)/ERC Grant Agreement 
n. 306476. GC is supported by Research Fellowships of the Japan Society for the 
Promotion of Science (JSPS) for Young Scientists. NY is grateful for financial support by the Grants-in-Aid
by the Ministry of Education, Science and Culture of Japan (25287050).
We aknowledge financial support from PRIN MIUR 2010-2011, project ``The Chemical and dynamical Evolution 
of the Milky Way and Local Group Galaxies'', prot. 2010LY5N2T.


\begin{thebibliography}{}
\bibitem[{Andreazza \& Singh}{~1997}]{AS1997} Andreazza, C. M., Singh, P. D., 1997, \mnras, 287, 287
\bibitem[{Aoki et al.}{~2010}]{aok2010} Aoki, W., Beers, T. C., Honda, S., \& Carollo, D., 2010, \apjl, 723, L201
\bibitem[{Asplund et al.}{~2009}]{Asp2009} Asplund, M., Grevesse, N., Sauval, A. J., \& Scott, P., 2009, ARA\&A , 47, 481
\bibitem[{Beers \& Christlieb}{~2005}]{Bee2005} Beers, T. C. \& Christlieb, N., 2005, ARA\&A, 43, 531
\bibitem[{Babb \& Dalgarno}{~1995}]{Babb1995} Babb, J. F. \& Dalgarno, A., 1995, Phys. Rev. A, 51, 3021
\bibitem[{Bianchi \& Schneider}{~2007}]{Bia2007} Bianchi, S. \& Schneider, R., 2007, \mnras, 378, 973 
\bibitem[{Bromm \& Loeb}{~2003}]{Bro2003} Bromm, V. \& Loeb, A., 2003, Nature, 425, 812
\bibitem[{Caffau et al.}{~2011}]{Caf11} Caffau, E., Bonifacio, P., Francois, P., Sbordone, L., Monaco, L., Spite, M., Ludwig, H.G., Cayrel, R. et al., 2011, Nature, 477, 67
\bibitem[{Cayrel et al.}{~2004}]{Cay2004} Cayrel, R., Depagne, E., Spite, M., Hill, V., Spite, F., Francois, P., Plez, B., Beers, T. et al., 2004, A\&A, 416, 1117
\bibitem[{Campbell et al.}{~2010}]{Cam2010} Campbell, S. W., Lugaro, M.,\& Karakas, A. I., 2010, A\&A, 522, L6
\bibitem[{Cherchneff \& Dwek}{~2009}]{CD2009} Cherchneff, I.,\& Dwek, E., 2009, \apj, 703, 642
\bibitem[{Cherchneff \& Dwek}{~2010}]{CD2010} Cherchneff, I.,\& Dwek, E., 2010, \apj, 713, 1 
\bibitem[{Chiaki et al.}{~2013}]{Chi2013} Chiaki, G., Nozawa, T., \& Yoshida, N., 2013, \apjl, 765, L3
\bibitem[{Chiaki et al.}{~2014}]{Chi2014} Chiaki, G., Schneider, R., Nozawa, T., Omukai, K., Limongi, M., Yoshida, N., Chieffi, A., 2014, \mnras, 439, 3121
\bibitem[{Dalgarno et al.}{~1990}]{Dal1990} Dalgarno, A., Du, M. L., \& You, J. H., 1990, ApJ, 349, 675
\bibitem[{Frebel et al.}{~2007}]{Fre2007} Frebel, A., Johnson, J. L. \& Bromm, V., 2007, \mnras, 380, L40
\bibitem[{Frebel \& Norris}{~2013}]{Fre2013} Frebel, A. \& Norris, J. E., 2013, Planets, Stars and Stellar Systems, Springer
\bibitem[{Heger \& Woosley}{~2010}]{Heg2010} Heger, A., \& Woosley, S.E., 2010, \apj, 724, 341
\bibitem{}Hirano, S., Hosokawa, T.,  Yoshida, N., Umeda, Y., Omukai, K., Chiaki, G., and Yorke, H. W., 2014, \apj, 781, 60
\bibitem[{Hirschi}{~2007}]{Hir2007} Hirschi, R., 2007, A\&A, 461, 571
\bibitem[\protect\citeauthoryear{Ji et al.}{~2014}]{Ji2014} Ji, A. P., Frebel, A., \& Bromm, V., 2014, \apj, 782, 95
\bibitem[{Joggerst et al.}{~2009}]{Jog2009} Joggerst, C. C., Woosley, S. E., \& Heger, A., 2009, \apj, 693, 1780
\bibitem[{Keller et al.}{~2014}]{Kel2014} Keller, S. C., Bessel, M. S., Frebel, A., Casey, A.R., Asplund, M., Jacobson, H. R., Lind, K., Norris, J.E. et al., 2014, Nature, 506, 463   
\bibitem[{Limongi \& Chieffi}{~2012}]{Lim2012} Limongi, M. \& Chieffi, A. 2012, \apjs, 199, 38L
\bibitem[{Limongi \& Chieffi}{~2006}]{Lim2006} Limongi, M. \& Chieffi, A. 2006, \apj,  647, 483
\bibitem[{Limongi \& Chieffi}{~2003}]{Lim2003} Limongi, M. \& Chieffi, A. 2003, \apj,  592, 404
\bibitem[{Lucatello et al.}{~2005}]{Luc2005}Lucatello, S., Tsangarides, S., Beers, T. C., Carretta, E., Gratton, R. G., Ryan, S.G., 2005, ApJ, 625, 825
\bibitem[{Madau et al.}{~2001}]{Mad2001} Madau, P., Ferrara, A., Rees, M. J., 2001, \apj, 555, 92 
\bibitem[{Marassi et al.}{~2014a}]{Mar2014a} Marassi, S., Schneider, R., Limongi, M., Chieffi, A. 2014, Proceedings of Science, 
{\it The Life Cycle of Dust in the Universe: Observations, Theory, and Laboratory Experiments - LCDU 2013, 18-22 November 2013 Taipei, Taiwan}
\bibitem[{Marassi et al.}{~2014b}]{Mar2014b}Marassi, S., Schneider, R., Limongi, M., Chieffi, A. 2014, in preparation
\bibitem[{Masseron et al.}{~2010}]{Mas2010} Masseron, T., Johnson, J. A., Plez, B., Van Eck, S. Primas, F., Goriely, S., Jorissen, A., 2010, A\&A, 509, 93
\bibitem[{Meynet et al.}{~2006}]{Mey2006} Meynet, G., Ekstr\"{o}m, S., \& Maeder, A. 2006, A\&A, 447, 623
\bibitem[{Meynet et al.}{~2010}]{Mey2010} Meynet, et al. 2010, A\&A, 521, A30
\bibitem[{Norris et al.}{~2013}]{Nor2013} Norris, J. E. et al. 2013, \apj, 762, 28
\bibitem[{Nozawa \& Kozasa}{~2003}]{NK2003} Nozawa, T. \& Kozasa, T. 2003, \apj 598, 785 
\bibitem[{Nozawa et al.}{~2007}] {Noz2007} Nozawa, T., Kozasa, T., Habe, A., Dwek, E., Umeda, H., Tominaga, N., Maeda, K., Nomoto, K. 2007, ApJ, 666, 955
\bibitem[{Nozawa et al.}{~2010}]{Noz2010} Nozawa, T., Kozasa, T., Tominaga, N., Maeda, K., Umeda, H., Nomoto, K., Krause, O., 2010, 713, 356
\bibitem[{Nozawa et al.}{~2012}]{Noz2012} Nozawa, T., Kozasa, T., Nomoto, K. 2012, \apjl, 756, 35L
\bibitem[{Omukai}{~2000}]{Omu2000} Omukai, K., 2000, \apj, 534, 809
\bibitem[{Omukai et al.}{~2005}]{Omu2005} Omukai, K., Tsuribe, T., Schneider, R., Ferrara, A., 2005, \apj, 626, 627
\bibitem[{Paquette \& Nuth}{~2011}]{PN2011} Paquette, J. A., \& JNuth~III, J.A. 2011, \apjl, 737, L6
\bibitem[{Sarangi \& Cherchneff}{~2013}]{Sar2013}Sarangi, A. \& Cherchneff, I. 2013, \apj, 776, 107 
\bibitem[{Schneider et al.}{~2003}]{Sch2003} Schneider, R., Ferrara, A., Salvaterra, R., Omukai, K., \& Bromm, V. 2003, Nature, 422, 869
\bibitem[{Schneider et al.}{~2004}]{Sch2004} Schneider, R., Ferrara, A., Salvaterra R. 2004, \mnras, 351, 1379
\bibitem[{Schneider et al.}{~2010}]{Sch2010} Schneider, R. \& Omukai, K. 2010, \mnras, 402, 429
\bibitem[{Schneider et al.}{~2012a}]{Sch2012a} Schneider, R., Omukai, K., Bianchi, S., Valiante, R., 2012a, \mnras, 419, 1566
\bibitem[{Schneider et al.}{~2012b}]{Sch2012b} Schneider, R., Omukai, K., Limongi M., Ferrara A., Salvaterra R., Chieffi A., Bianchi, S., 2012b, \mnras, 423, L60
\bibitem[{Schneider et al.}{~2014}]{Sch2014} Schneider, R., Valiante, R., Ventura, P., dell'Agli, F., Di Criscienzo, M., Hirashita, H., Kemper, F. 2014, \mnras, in press
\bibitem[{Silvia et al.}{~2010}]{Sil2010} Silvia, D. W., Smith, B., D., \& Shull, J. M. 2010, ApJ, 715, 1575
\bibitem[{Silvia et al.}{~2012}]{Sil2012} Silvia, D. W., Smith, B., D., \& Shull, J. M. 2012, ApJ, 748, 12
\bibitem[{Suda et al.}{~2004}]{Sud2004} Suda, T., Aikawa, M., Machida, M. N., Fujimoto, M. Y., Iben, I. J. 2004, \apj, 611, 476
\bibitem[{Tanaka \& Omukai}{~2014}]{Tan2014} Tanaka, K. E. I.\& Omukai, K., 2014, \mnras, 439, 1884
\bibitem[{Todini \& Ferrara}{~2001}]{Tod2001} Todini, P. \& Ferrara, A. 2001, \mnras, 325, 726
\bibitem[{Tominaga et al.}{~2007}]{Tom2007} Tominaga, N., Umeda, H., \& Nomoto, K. 2007, \apj, 660, 516
\bibitem[{Truelove \& McKee}{~1999}]{True1999} Truelove, J. K. \& McKee, C. F. 1999, ApJ, 120, 299 
\bibitem[{Umeda \& Nomoto}{~2002}]{Ume2002} Umeda, H. \& Nomoto, K. 2002, \apj, 565, 385
\bibitem[{Umeda \& Nomoto}{~2003}]{Ume2003} Umeda, H. \& Nomoto, K. 2003, Nature, 422, 871
\bibitem[{Umist database}{~2012}]{Umist} UMIST datatbase for astrochemistry 2012, http://udfa.ajmarkwick.net
\bibitem[{Woosley}{1988}]{Woo1988} Woosley, S. E., 1988, \apj, 330, 218
\bibitem[{Woosley et al.}]{Woo1989} Woosley, S.E., Pinto, P. A., Hartmann, D., 1989, \apj, 346, 395
\bibitem[{Valiante \& Schneider}{~2014}]{Val2014} Valiante, R. \& Schneider, R. 2014, Proceedings of Science, 
{\it The Life Cycle of Dust in the Universe: Observations, Theory, and Laboratory Experiments - LCDU 2013, 18-22 November 2013 Taipei, Taiwan}
\bibitem[{Yong et al.}{~2013a}]{Yon2013a} Yong, D., Norris, J. E., Bessell, M. S., Christlieb, N., Asplund, M., Beers, T.C., Barklem, P. S., Frebel, A. et al., 2013a, \apj, 762, 26
\bibitem[{Yong et al.}{~2013b}]{Yon2013b}Yong, D., Norris, J. E., Bessell, M. S., Christlieb, N., Asplund, M., Beers, T.C., Barklem, P. S., Frebel, A. et al., 2013b, \apj, 762, 27
\end{thebibliography}
\end{document}